%% file: main.tex
\documentclass[
pra,aps,
reprint,
a4paper,
amsmath,amssymb,amsbsy,
floatfix
]{revtex4-1}
\usepackage{etoolbox}
	\newtoggle{arXiv}\toggletrue{arXiv}
	\newtoggle{bibrun}%\toggletrue{bibrun}					
\usepackage{newtxtext, newtxmath}
\usepackage{graphicx}
\usepackage{color, calc}
\usepackage{array, booktabs}
	\newcolumntype{L}{>{$}l<{$}}
	\newcolumntype{C}{>{$}c<{$}}
	\newcolumntype{R}{>{$}r<{$}}
\usepackage{tikz, tikz-3dplot}
	\usetikzlibrary{arrows, arrows.meta, calc, positioning}
	\usetikzlibrary{decorations.pathmorphing}	
	\tikzset{>=Stealth}
\usepackage{pgfplots}
	\pgfplotsset{compat=1.14, small, scale only axis}
\usepackage{siunitx, physics}
	\newcommand{\I}{\ensuremath{\mathrm{i}}}
	\newcommand{\Exp}[1]{\mathrm{e}^{#1}}
\usepackage{enumitem}
\setlist[description]{labelindent=0pt, leftmargin=\parindent, font=\normalfont\itshape}

\usepackage{hyperref}

\begin{document}

%%%%%%%%%%%%%%%%%%%%%%%%%%%%%%%%%%%%%%%%%%%%%%%%%%%%%%%%%%%%%%%%%%%

\title{Sustained state-independent quantum contextual correlations from a single ion}

\author{F. M. Leupold}
\author{M. Malinowski}
\author{C. Zhang}
\author{V. Negnevitsky}
\author{J. Alonso}\email{alonso@phys.ethz.ch}
\author{J. P. Home}\email{jhome@phys.ethz.ch}
\affiliation{Institute for Quantum Electronics, ETH Z\"urich, Otto-Stern-Weg 1, 8093 Z\"urich, Switzerland}
\author{A. Cabello}
\affiliation{Departamento de F\'isica Aplicada II, Universidad de Sevilla, 41012 Sevilla, Spain}

%%%%%%%%%%%%%%%%%%%%%%%%%%%%%%%%%%%%%%%%%%%%%%%%%%%%%%%%%%%%%%%%%%%

\begin{abstract}
We use a single trapped-ion qutrit to demonstrate the violation of input-state-independent non-contextuality inequalities using a sequence of randomly chosen quantum non-demolition projective measurements. We concatenate 53 million sequential measurements of 13 observables, and violate an optimal non-contextual bound by 214 standard deviations. We use the same dataset to characterize imperfections including signaling and repeatability of the measurements. The experimental sequence was generated in real time with a quantum random number generator integrated into our control system to select the subsequent observable with a latency below $\SI{50}{\micro\second}$, which can be used to constrain hidden-variable models that might describe our results. The state-recycling experimental procedure is resilient to noise, self-correcting and independent of the qutrit state, substantiating the fact that the contextual nature of quantum physics is connected to measurements as opposed to designated states. The use of extended sequences of quantum non-demolition measurements finds applications in the fields of sensing and quantum information. 
\end{abstract}

\maketitle

%%%%%%%%%%%%%%%%%%%%%%%%%%%%%%%%%%%%%%%%%%%%%%%%%%%%%%%%%%%%%%%%%%%
%% Introduction

Two measurements are said to be compatible when the outcome statistics of each of them individually is independent of whether the other is carried out or not. In classical theories, outcomes of measurements are consistent with each measurement result having a pre-existing value, independent of which other compatible measurements are performed. However, correlations between the outcomes of compatible observables in Quantum Mechanics (QM) can be stronger than in classical theories. This feature, which is known as contextuality, has been linked to the power of quantum computation \cite{14Howard,16Bermejo,17Bermejo} and its most famous manifestation is Bell non-locality \cite{64Bell}. In this sense, the violation of a Bell inequality demonstrates contextuality. However, non-locality requires composite systems in entangled states. A more general result is that of Kochen and Specker \cite{67Kochen}, who showed that any state of any quantum system in a Hilbert space of dimension greater than 2 can be used to reveal contextuality.

In a similar sense to a Bell-inequality, the contextuality of QM can be shown through the violation of a number of inequalities, which have been derived for systems of various Hilbert space dimension. Such inequalities can be split into those which are violated for a given input state \cite{69Clauser,08Klyachko}, and those for which the violation is input-state independent \cite{08Cabello,12Yu}. We will refer to the latter as State-Independent-Contextuality (SIC) tests. SIC tests have been performed using a number of systems \cite{09Kirchmair2,09Amselem,10Moussa,12Zu,13Zhang,13DAmbrosio,13Huang,14Canas,14Canas2}, but thus far they all used the following approach: i) prepare an input states, ii) measure multiple observables. This was repeated for each of a finite number of input states, and using all combinations of observables required for the test. Measurements on each observable can either be carried out simultaneously or sequentially \cite{09Guhne}, with the sequential approach being the most popular.

An alternative proposal \cite{16Wajs} is to perform a SIC test using sequences of ideal Quantum Non-Demolition (QND) projective measurements (which in the context of general probabilistic theories are known as sharp measurements \cite{14Chiribella}). Each measurement is performed on the state into which the system was projected by the previous measurement. When executed in this manner, contextuality tests intrinsically stabilize the generation of quantum correlations and are self-correcting, which can be used to generate and certify continuous strings of random numbers \cite{ThColbeck,13Um}.

In this Letter, we demonstrate SIC sustainable in time using state-recycling over a sequence of 53 million measurements. To that end we have adopted: i) the simplest system featuring SIC, a three-level quantum system or qutrit \cite{66Bell,67Kochen}, ii) the smallest set of elementary quantum measurements needed for SIC, namely, the Yu-Oh set with 13 observables \cite{12Yu,15Cabello,16Cabello}, and iii) the original Yu-Oh and an optimal witness of SIC \cite{12Kleinmann}. Our results violate the bounds imposed by non-contextual hidden-variable models. We use a commercial Quantum Random Number Generator (QRNG) to create the sequence of measured observables in real time. This places constraints on contextual hidden-variable models attempting to explain our results, which must cover the behavior not only of the qutrit but also of the QRNG \cite{88Peres}. We quantify the sharpness and compatibility of our measurements by extracting high-order correlators from the dataset.

%%%%%%%%%%%%%%%%%%%%%%%%%%%%%%%%%%%%%%%%%%%%%%%%%%%%%%%%%%%%%%%%%%%
%% YuOh Scheme

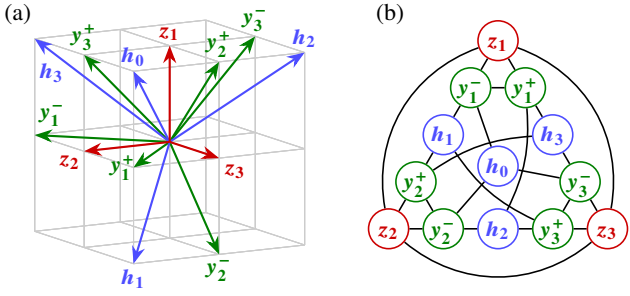
\begin{figure}
\input{fig_RayStructure}
\caption{(color online) Observables and compatibility relations between the observables for the Yu-Oh set. (a) The 13 rays are represented by vectors in a three-dimensional real Hilbert space. Their directional components are listed in TABLE~\ref{tab:RayAngles}. (b) The orthogonality relationships between the rays determine a graph with 13 vertices and 24 edges between compatible rays.}
\label{fig:YuOhRays}
\end{figure}

The 13 dichotomic (``yes-no'') observables or ``rays'' in the Yu-Oh set \cite{12Yu} are of the form $A_v=I-2P_v$, where $I$ is the identity, $P_v$ is the normalized projection operator onto a vector $\ket*{v}=a\ket*{0}+b\ket*{1}+c\ket*{2}$, and $\Bqty{\ket*{0},\ket*{1},\ket*{2}}$ form a qutrit basis. Since the eigenvalues of $P_v$ are 0 and 1, ray measurements result in values $+1$ and $-1$. The 13 vectors $\ket*{v}$ with real-valued coefficients $\qty(a,b,c)$ are defined by points on the surface of a $3\times 3$ cube in a three-dimensional Hilbert space (FIG.~\ref{fig:YuOhRays}a, TABLE~\ref{tab:RayAngles}). Two rays are compatible if the corresponding vectors are orthogonal. This can be visualized in an orthogonality graph (FIG.~\ref{fig:YuOhRays}b) by drawing all vectors from the set $V=\qty\big{y_k^\sigma,h_\alpha,z_k|k=1,2,3;\sigma=\pm;\alpha=0,1,2,3}$ as vertices and linking vertices of compatible rays. In this notation, $z_k$ are the basis states, $y_k$ are superpositions of two basis states, and $h_k$ are superpositions of all three. In total, there are 24 edges in the graph, representing the 24 compatible pairs $(u,v)\in E$ with $P_uP_v=0$ (each edge is counted only once). 

Besides the original Yu-Oh witness \cite{12Yu}
\begin{subequations}	\label{eq:SICWitnesses}
\begin{align}	
\expval*{\chi_\text{YO}}=\sum_{v\in V} \expval*{A_v} - \sum_{(u,v)\in E} \frac{1}{2}\,\expval*{A_u A_v},
\end{align}
we use the optimal SIC witness opt3 for which the QM and classical predictions differ maximally \cite{12Kleinmann}
\begin{align}
\expval*{\chi_\text{opt3}}=&\sum_{v\in V_h}2\,\expval*{A_v}+\sum_{v\in V\setminus V_h} \expval*{A_v}	\notag\\
		&-\sum_{(u,v)\in E\setminus C_2}2\,\expval*{A_u A_v}-\sum_{(u,v)\in C_2} \expval*{A_u A_v}	\notag\\
		&-\sum_{(u,v,w)\in C_3}3\,\expval*{A_u A_v A_w}.
\end{align}
\end{subequations}
Here $V_h=\qty\big{h_\alpha}$, $C_2=\qty\big{\pqty\big{z_k,y_k^+},\pqty\big{z_k,y_k^-},\pqty\big{y_k^+,y_k^-}}$ and $C_3=\qty\big{\pqty\big{z_k,y_k^+,y_k^-}}$, with indices $k$ and $\alpha$ running as for $V$. A necessary condition for a set of correlations to be non-contextual is
\begin{align}	\label{eq:NoncontextualInequalities}
\expval*{\chi_\text{YO}}\leq 8	\qq{and}
\expval*{\chi_\text{opt3}}\leq 25,
\end{align}
and any violation of these inequalities demonstrates contextuality. The prediction of quantum theory is that, for any qutrit state and under ideal conditions,
\begin{align}
\expval*{\chi_\text{YO}}=\frac{25}{3}\approx 8.333	\qq{and}
\expval*{\chi_\text{opt3}}=\frac{83}{3}\approx 27.667.
\end{align}

\begin{table}[hbtp]
\centering
\caption[]{Definition and experimental parameters for the vectors $v\in V$ in the Yu-Oh set. The coefficients $(a,b,c)$ give the directions of the rays in the real-valued three-dimensional Hilbert space (FIG.~\ref{fig:YuOhRays}). In the experiment, rays are rotated onto the measurement axis (along $z_1$) by applying the coherent rotations in Equations~\eqref{eq:RotationOperators} using the angles $\theta^{(1)}_v$, $\phi^{(1)}_v$, $\theta^{(2)}_v$, $\phi^{(2)}_v$ (see also FIG.~\ref{fig:MeasurementSequence}). The last column shows the corresponding bit sequence from the QRNG (see text for details). If the QRNG delivers a bit sequence not present in this table, it is discarded and a new one is read in. Shorthand notations $\bar{1}=-1$ and  $\theta^{(2)}_{h}=2\arctan\pqty\big{1/\sqrt{2}}$ were used.}
\label{tab:RayAngles}%
\newcommand{\nop}[1]{0}
\begin{ruledtabular}\renewcommand{\arraystretch}{1.2}
\begin{tabular}{LRRRRRc}
	v     &       (a,b,c) & \theta^{(1)}_v & \phi^{(1)}_v &   \theta^{(2)}_v & \phi^{(2)}_v & QRNG \\\colrule
	y_1^- & (0,1,\bar{1}) &            \pi &       3\pi/2 &            \pi/2 &        \pi/2 & 0001 \\
	y_2^- & (\bar{1},0,1) &              0 & \nop{3\pi/2} &           3\pi/2 &       3\pi/2 & 0010 \\
	y_3^- & (1,\bar{1},0) &          \pi/2 &        \pi/2 &                0 & \nop{3\pi/2} & 0011 \\
	y_1^+ &       (0,1,1) &            \pi &       3\pi/2 &            \pi/2 &       3\pi/2 & 0100 \\
	y_2^+ &       (1,0,1) &              0 & \nop{3\pi/2} &            \pi/2 &       3\pi/2 & 0101 \\
	y_3^+ &       (1,1,0) &          \pi/2 &       3\pi/2 &                0 & \nop{3\pi/2} & 0110 \\
	h_1   & (\bar{1},1,1) &         3\pi/2 &       3\pi/2 & \theta^{(2)}_{h} &       3\pi/2 & 0111 \\
	h_2   & (1,\bar{1},1) &          \pi/2 &        \pi/2 & \theta^{(2)}_{h} &       3\pi/2 & 1000 \\
	h_3   & (1,1,\bar{1}) &          \pi/2 &       3\pi/2 & \theta^{(2)}_{h} &        \pi/2 & 1001 \\
	h_0   &       (1,1,1) &          \pi/2 &       3\pi/2 & \theta^{(2)}_{h} &       3\pi/2 & 1010 \\
	z_1   &       (1,0,0) &              0 & \nop{3\pi/2} &                0 & \nop{3\pi/2} & 1011 \\
	z_2   &       (0,1,0) &            \pi &       3\pi/2 &                0 & \nop{3\pi/2} & 1100 \\
	z_3   &       (0,0,1) &              0 & \nop{3\pi/2} &              \pi &       3\pi/2 & 1101
\end{tabular}
\end{ruledtabular}
\end{table}%

%%%%%%%%%%%%%%%%%%%%%%%%%%%%%%%%%%%%%%%%%%%%%%%%%%%%%%%%%%%%%%%%%%%
%% Experimental realization

\begin{figure}
\input{fig_LevelDiagram}
\caption{(color online) Energy level diagram of the ${}^{40}\text{Ca}^+$ ion. Qutrit states $\ket*{0}$, $\ket*{1}$, and $\ket*{2}$ are encoded in the highlighted fine-structure levels. Coherent rotations between them are achieved with laser pulses at \SI{729}{nm}. Fluorescence measurements using an excitation laser at \SI{397}{nm} project the qutrit state into either $\ket*{0}$ (``bright'') or the $\ket*{1}$,$\ket*{2}$-manifold (``dark'').}
\label{fig:LevelDiagram}
\end{figure}
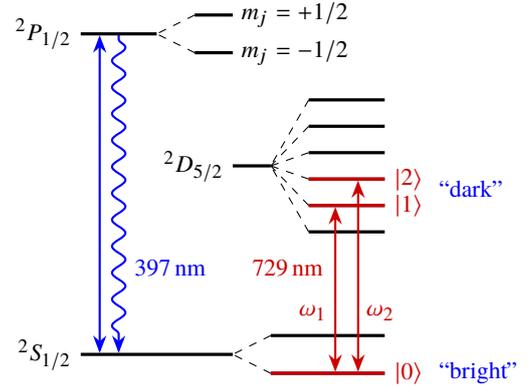

Our experimental platform to test these witnesses uses a single ${}^{40}\text{Ca}^+$ ion confined in a surface-electrode radio-frequency trap in the setup described in \cite{16Alonso}. The qutrit basis states are represented by three fine-structure levels in a ${}^{40}\text{Ca}^+$ ion: $\ket*{0}=\ket*{S_{1/2}(m_j=-1/2)}$ in the ground-state manifold, and $\ket*{1}=\ket*{D_{5/2}(m_j=-3/2)}$ and $\ket*{2}=\ket*{D_{5/2}(m_j=-1/2)}$ in the metastable $D_{5/2}$ manifold (FIG.~\ref{fig:LevelDiagram}). The two metastable states have a Zeeman-shifted energy difference $\hbar(\omega_2-\omega_1)=(2\pi\hbar)\,\SI{6.47}{\mega\hertz}$ in an external magnetic field of $B=\SI{0.385}{\milli\tesla}$.

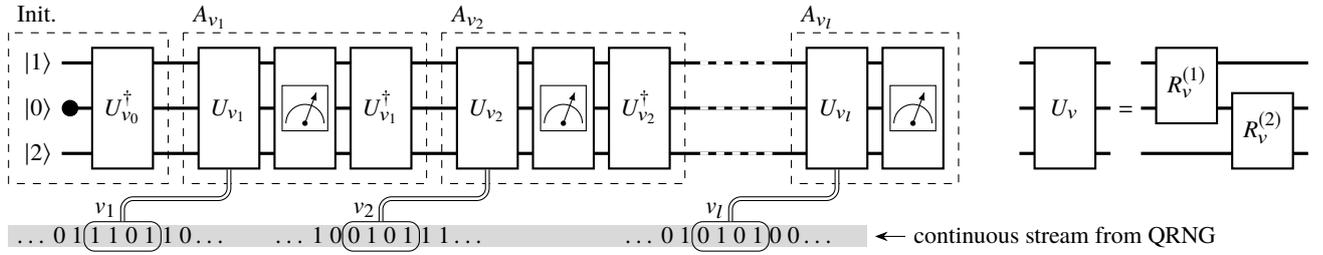
\begin{figure*}
\input{fig_MeasurementScheme}
\caption{Sequential measurement scheme. A subsequence starts by initializing the ion state to $\ket*{0}$ and rotating it to the last ray from the previous subsequence, $v_0$. Every following ray measurement $A_v$ then consists of a unitary transformation $U_v$ rotating the ray $v$ onto $\ket*{0}$, a projective measurement, and the back rotation $U_v^\dagger$. The unitary transformations $U_v=R^{(2)}_v\,R^{(1)}_v$ are realized by coherent driving on the transitions between $\ket*{0}$ and $\ket*{1}$ , $R^{(1)}_v=R^{(1)}\big(\theta^{(1)}_v,\phi^{(1)}_v\big)$, and between $\ket*{0}$ and $\ket*{2}$, $R^{(2)}_v=R^{(2)}\big(\theta^{(2)}_v,\phi^{(2)}_v\big)$. Subsequent measurement rays are determined by bit sequences from a QRNG, which are created after performing the respective previous projective measurement.}
\label{fig:MeasurementSequence}
\end{figure*}

Every experimental sequence starts with \SI{500}{\micro\second} of Doppler cooling using a \SI{397}{\nano\meter} laser red-detuned approximately half a natural linewidth from resonance with the cycling transition between the $S_{1/2}$ and $P_{1/2}$ manifolds, and with close to one saturation intensity \cite{SuppMat,16Alonso}. This is followed by \SI{10}{\micro\second} of optical pumping to initialize the qutrit to the $\ket*{0}$ state. Subsequently, measurements of the observables $\{A_v\}$ are performed, which consist of coherent rotations between the qutrit states and projective measurements. Coherent rotations are achieved using \SI{729}{\nano\meter} laser pulses resonant with the transitions between $\ket*{0}$ and $\ket*{1}$ (at $\omega_1$), and between $\ket*{0}$ and $\ket*{2}$ (at $\omega_2$). Matrix representations of the rotations in the Hilbert space spanned by the basis $\qty\big{\ket*{0},\ket*{1},\ket*{2}}$ are given by
\begin{subequations}	\label{eq:RotationOperators}
\begin{align}
	R^{(1)}(\theta,\phi)&=\left(\begin{array}{ccc}
		\cos(\frac{\theta}{2}) & -\I \Exp{-\I\phi} \sin(\frac{\theta}{2}) & 0 \\[0.5ex]
		-\I \Exp{\I\phi} \sin(\frac{\theta}{2}) & \cos(\frac{\theta}{2}) & 0 \\[0.5ex]
		0 & 0 & 1
	\end{array}\right),	\\
	R^{(2)}(\theta,\phi)&=\left(\begin{array}{ccc}
		\cos(\frac{\theta}{2}) & 0 & -\I \Exp{-\I\phi} \sin(\frac{\theta}{2}) \\[0.5ex]
		0 & 1 & 0 \\[0.5ex]
		-\I \Exp{\I\phi} \sin(\frac{\theta}{2}) & 0 & \cos(\frac{\theta}{2})
	\end{array}\right).
\end{align}
\end{subequations}
The angles $\theta$ and $\phi$ for a certain rotation (TABLE~\ref{tab:RayAngles}) are controlled via the duration and phase of the corresponding laser pulse using an acousto-optic modulator. Projective measurements are realized by illuminating the ion for \SI{160}{\micro\second} with the same settings used for Doppler cooling \cite{SuppMat}. If photons are scattered, the qutrit state is projected onto $\ket*{0}$ (``bright state''); if not, the qutrit is projected onto the $D_{5/2}$ manifold (``dark states''), preserving the coherence between $\ket*{1}$ and $\ket*{2}$ (FIG.~\ref{fig:LevelDiagram}). For the bright / dark states, we register on average \num{18.8} / \num{0.7} photons through a high-numerical aperture objective on a photomultiplier tube. Thresholding single-shot photon counts at \num{5.5} for the \SI{160}{\micro\second} detection window allows us to distinguish bright from dark states with an estimated mean detection-error of $<\num{2e-4}$ \cite{SuppMat}.

Testing the SIC inequalities on the Yu-Oh set \cite{12Yu} requires projective measurements along all 13 rays (FIG.~\ref{fig:YuOhRays}). By design, the fluorescence detection projects onto either the qutrit state $\ket*{0}$ itself, i.e. the $z_1$ ray, or the plane orthogonal to it, spanned by $\ket*{1}$ and $\ket*{2}$. For any other observable $A_v$, we apply first a unitary rotation $U_v=R^{(2)}\pqty\big{\theta^{(2)}_v,\phi^{(2)}_v}\,R^{(1)}\pqty\big{\theta^{(1)}_v,\phi^{(1)}_v}$, which rotates $v$ onto $z_1$, then fluorescence detection (followed by optical pumping of the $S_{1/2}$ population to $\ket*{0}$), and finally the reverse rotation $U_v^\dagger$ (FIG.~\ref{fig:MeasurementSequence}). Every measurement of an observable is thus uniquely determined by $v$ and is independent of the context.

Ideally, we would perform a single long series of measurements of randomly chosen observables. In practice, we interrupt the sequence to save collected data and periodically calibrate laser frequencies and pulse times. To sustain the sequence, we take subsequences containing a minimum of 1,000 measurements, which we interrupt when the last detection projected the qutrit onto $\ket*{0}$. The next subsequence then starts by initializing the qutrit to $\ket*{0}$ and applying the rotation $U_{v_0}^\dagger$, with $v_0=v_l$ the last ray from the previous sequence. In this way, all performed measurement sequences can be concatenated up to the 53 million in the present dataset \cite{SuppMat}. 

We randomize the sequence of measured observables using a QRNG (model Quantis from ID Quantique SA). It delivers a constant stream of random bits, from which we take groups of four and assign rays $v$ to them (TABLE~\ref{tab:RayAngles}). The random bits for an observable are created after the detection event of the previous observable (FIG.~\ref{fig:MeasurementSequence}). In this way, if we acknowledge the randomness of the QRNG, we prevent a hypothetically conspiring ion from knowing what the context of a measurement will be \cite{88Peres}. Everything from the QRNG output to the pulse sequence programmed in the computer-control system is updated in real time within a \SI{50}{\micro\second} time window between unitary rotations.

In a typical sequence of 1 million measurements, we observe between two and five subsequences containing more than \num{55} dark measurements in a row. In a random sequence of 55 ideal measurements, we would, however, only expect such a set to occur with a probability of $(2/3)^{55}\approx\num{2e-10}$, which corresponds to a \SI{1}{\percent} probability for it to appear once in the full set of 53 million measurements. We attribute this anomalous effect to off-resonant leakage into the states $\ket*{D_{5/2},m_j=-5/2}$ and  $\ket*{D_{5/2},m_j=+1/2}$, which are long-lived dark states outside our computational Hilbert space \cite{SuppMat}. The control system for the experiment spots these events in real time and breaks, purging the subsequence and starting a new subsequence from the same $v_0$ as was used for the purged subsequence.

%%%%%%%%%%%%%%%%%%%%%%%%%%%%%%%%%%%%%%%%%%%%%%%%%%%%%%%%%%%%%%%%%%%
%% Data analysis and results

\begin{figure}
\input{fig_CorrelatorResults}
\caption{Experimental results for expectation values that enter the SIC witnesses in Equations~\eqref{eq:SICWitnesses} (see text for details on their calculation). Error bars reflect shot noise; dashed lines represent values predicted by quantum mechanics.}
\label{fig:Correlators}
\end{figure}
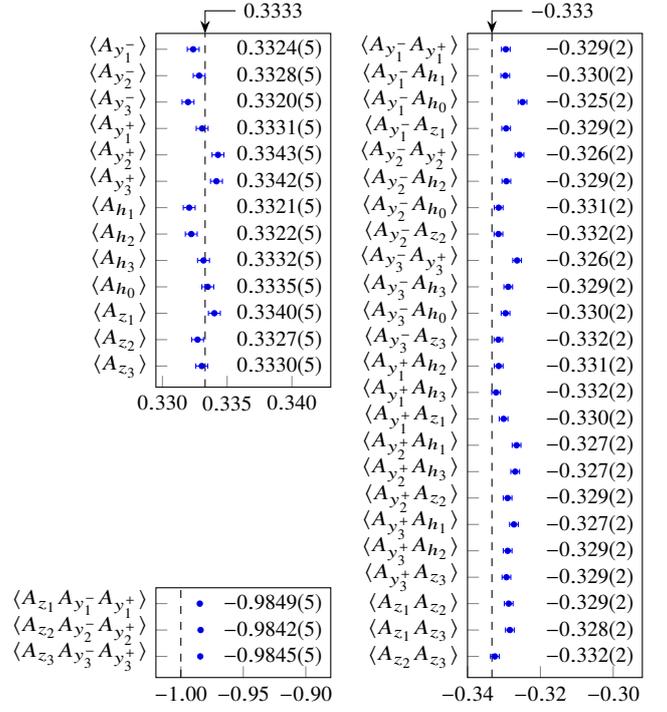

Every data point measured for an observable $A_v$ consists of the measurement ray $v$ and an outcome $a=\pm 1$. From the full data set, we collect the numbers $N(A_v{=}a_1)$, $N(A_u{=}a_1,A_v{=}a_2)$, and $N(A_u{=}a_1,A_v{=}a_2,A_w{=}a_3)$, where $A_u$, $A_v$, and $A_w$ are successive measurements in that order, for all $u,v,w \in V$ and all $a_1,a_2,a_3 \in \Bqty{1,-1}$.
Based on these numbers, we compute the expectation values
\begin{subequations}
\begin{align}
\expval*{A_v}&=\frac{\sum_{a_1} a_1 N(A_v{=}a_1)}
		{\sum_{a_1} N(A_v{=}a_1)},	\\
\expval*{A_u A_v}&=\frac{\sum'_{a_1,a_2} a_1 a_2 N(A_u{=}a_1,A_v{=}a_2)}
		{\sum'_{a_1,a_2} N(A_u{=}a_1,A_v{=}a_2)}, \\
\expval*{A_u A_v A_w}&=\frac{\sum'_{a_1,a_2,a_3} a_1 a_2 a_3 N(A_u{=}a_1,A_v{=}a_2,A_w{=}a_3)}
{\sum'_{a_1,a_2,a_3} N(A_u{=}a_1,A_v{=}a_2,A_w{=}a_3)},
\end{align}
\end{subequations}
where $\sum'$ additionally sums over all permutations of the argument list of $N$, i.e. the measurement order. Substituting the obtained values (FIG.~\ref{fig:Correlators}) into the SIC witnesses in Equations~\eqref{eq:SICWitnesses}, we find
\begin{align}
% Threshold at 4.5:
%	\expval*{\chi_\text{YO}}=\num{8.266(4)}	\qq{and} \expval*{\chi_\text{opt3}}=\num{27.321(11)}.
% Threshold at 5.5:
	\expval*{\chi_\text{YO}}=\num{8.279(4)}	\qq{and} \expval*{\chi_\text{opt3}}=\num{27.357(11)}.
\end{align}
Our results thus violate Inequalities~\eqref{eq:NoncontextualInequalities} by 69 and 214 standard deviations. These deviations are solely based on statistical uncertainties, which are small due to the large number of measurements in the complete dataset. We, however, believe that significance of these violations should be penalized according to experimental imperfections and systematic errors \cite{SuppMat}, and elaborate on this issue below.

Our dataset additionally allows for evaluation of the SIC witnesses in Eqs.~(\ref{eq:SICWitnesses}) based on the ``standard approach'', where measurements are repeatedly performed on specifically prepared states of the system. For this, we calculate the averages conditioned on a preceding projection onto one of the states $i \in V$. We do this for all 13 input states and observe violations of the SIC inequalities by at least \num{15} and \num{43} standard deviations, respectively \cite{SuppMat}.

Inequalities~\eqref{eq:NoncontextualInequalities} are satisfied by any theory assuming non-contextuality and their violation indicates contextuality if certain underlying assumptions are satisfied. There are some such assumptions that are untestable, e.g., the assumption that observers have free will for choosing which measurement to make at any time (here implemented with a QRNG). Nevertheless, there are underlying assumptions that are (at least partially) testable. One is the assumption that measurements are sharp, i.e. they are minimally disturbing \cite{14Chiribella} and their outcomes are the same if performed repeatedly. Note that sharpness implies that measurements are repeatable even when other sharp compatible measurements are performed in between two successive realizations of the same measurement. In quantum theory, sharp measurements are represented by self-adjoint operators; the ``ideal measurements'' as defined by von Neumann \cite{32vonNeumann} are sharp measurements. While perfect sharpness can never be fulfilled in a real experiment, we find the repeatability of our measurements (including rotations and projections) to be above $\SI{99.6}{\percent}$
%TODO Add second number for interleaved measurements once this is in SM.
%, and above $\SI{98.9}{\percent}$ if a supposedly compatible measurement is interleaved 
\cite{SuppMat}. Another assumption is compatibility between the 24 pairs of observables in $E$ (FIG.~\ref{fig:YuOhRays}b). 
%TODO Add statement from PRL version?
While compatibility is among the untestable assumptions, one of its consequences are testable: given compatibility, there should be no context signaling in the data. We indeed find no signature of signaling backward in time, and attribute detectable, yet small traces of signaling forward in time to imperfect coherent rotations \cite{SuppMat}. The large number of measurements comprising our dataset render statistical uncertainties very small and we are able to resolve small systematic deviations from the ideal case in both these measures. We believe these should reflect on our results for the SIC witnesses (Eqs.~(\ref{eq:SICWitnesses})). But whereas there exist analytical methods to take into account such imperfections for non-contextuality inequalities for scenarios with cyclic systems in which dichotomic observables are measured in only two contexts \cite{15Kujala}, we are not aware of any standard method to account for these imperfections when evaluating SIC witnesses.

In \cite{SuppMat}, we characterize the quantum-vs-classical advantage of this experiment based on the fact that QM predictions for this system cannot be simulated with a classical trit as this would require a classical system with a substantially larger memory. Furthermore, we show that the compatibility structure between observables need not be assumed a priori, but can be inferred from the resulting statistics without invoking QM.

%%%%%%%%%%%%%%%%%%%%%%%%%%%%%%%%%%%%%%%%%%%%%%%%%%%%%%%%%%%%%%%%%%%
%% Conclusion and outlook

Beyond addressing fundamental aspects of QM, this work demonstrates a system capable of autonomously generating quantum operations, a feature desirable for a prospective quantum computer. The system concatenates hundreds of millions of coherent rotations and projective measurements, rather than repeating a finite sequence which starts with a pre-defined quantum state and consumes a resource at the end of the computation. Such long sequences of QND measurements are interesting in a range of areas including sensing and quantum computing \cite{13Haroche}. Furthermore, the methods presented in this paper might be generalized to multi-particle quantum systems, providing more powerful tests of fundamental physics \cite{12Cabello,16Zhan} and addressing the question of how to optimally generate, certify and make use of quantum contextual correlations.

%%%%%%%%%%%%%%%%%%%%%%%%%%%%%%%%%%%%%%%%%%%%%%%%%%%%%%%%%%%%%%%%%%%
%% Back matter

\begin{acknowledgments}
We thank IdQuantique for the QRNG, and Matt Grau for partial checking of our data analysis and comments on the manuscript. AC thanks Matthias Kleinmann for discussions, and Mile Gu and Zhen-Peng Xu for contributions to the memory estimation in \cite{SuppMat}. We acknowledge support from the Swiss National Science Foundation under grant no. 200021 134776, ETH Research Grant under grant no. ETH-18 12-2, and from the Swiss National Science Foundation through the National Centre of Competence in Research for Quantum Science and Technology (QSIT). The research is partly based upon work supported by the Office of the Director of National Intelligence (ODNI), Intelligence Advanced Research Projects Activity (IARPA), via the U.S. Army Research Office grant W911NF-16-1-0070. The views and conclusions contained herein are those of the authors and should not be interpreted as necessarily representing the official policies or endorsements, either expressed or implied, of the ODNI, IARPA, or the U.S. Government. The U.S. Government is authorized to reproduce and distribute reprints for Governmental purposes notwithstanding any copyright annotation thereon. Any opinions, findings, and conclusions or recommendations expressed in this material are those of the author(s) and do not necessarily reflect the view of the U.S. Army Research Office. 
AC acknowledges support from Project No.\ FIS2014-60843-P, ``Advanced Quantum Information'' (MINECO, Spain), with FEDER funds, the FQXi Large Grant ``The Observer Observed: A Bayesian Route to the Reconstruction of Quantum Theory,'' and the project ``Photonic Quantum Information'' (Knut and Alice Wallenberg Foundation, Sweden).
\end{acknowledgments}

\textbf{Author Contributions:} Experimental data were taken by FML, MM and JA, using an apparatus primarily built up by FML and JA, and with significant contributions from MM, CZ and VN. Data analysis was performed by FML, JA, JPH and AC. The paper was written by FML, JA, AC and JPH, with input from all authors. Experiments were conceived by AC, FML, JA and JPH.
The authors declare that they have no competing financial interests.

%% If main.bbl does not exist, untoggle arXiv and compile bibliographies for main.tex and supp.tex separately. While doing this, supp.tex must not cite any references that already occured in main.tex!
\iftoggle{arXiv}{\input{main.bbl}}{\bibliography{myrefs,myextrarefs}}

%%%%%%%%%%%%%%%%%%%%%%%%%%%%%%%%%%%%%%%%%%%%%%%%%%%%%%%%%%%%%%%%%%%
%% Include SI for arXiv version

\iftoggle{arXiv}{\input{supp_content}}{}

%%%%%%%%%%%%%%%%%%%%%%%%%%%%%%%%%%%%%%%%%%%%%%%%%%%%%%%%%%%%%%%%%%%
%% To dos

%TODO Add reference to supp: \cite{SuppMat}
% Authors should ensure that the journal article contains a reference in the reference list as follows:
% 	See Supplemental Material at [URL will be inserted by publisher] for [give brief description of material].

%TODO Check that counters in supp.tex are correct!

%TODO Add this point somewhere:
%Three results of the repeatability tests, namely
%	P_{z_1}^\text{(rep)}=\num{0.00046(4)},
%	P_{z_2}^\text{(rep)}=\num{0.00144(7)}, and
%	P_{z_3}^\text{(rep)}=\num{0.00153(7)},
%put lower bounds on our detection (z_1) and pulse fidelities (z_2 and z_3).

\end{document}

%% file: fig_RayStructure.tex
\tdplotsetmaincoords{79}{150}
\begin{tikzpicture}[cube line/.style={black!20},
z ray/.style={->, thick, red!80!black},
y ray/.style={->, thick, green!50!black},
h ray/.style={->, thick, blue!70},
every node/.style={font=\footnotesize},
vertex/.style={circle, draw, semithick, fill=white, inner sep=0pt, minimum size=15pt},
to/.style={semithick}]
\begin{scope}[scale=1.3, tdplot_main_coords]
\foreach \x in {-1,0,1}
\foreach \y in {-1,0,1}
\foreach \z in {-1,0,1}{
	\draw[cube line] (\x,-1,\z) -- (\x,1,\z);
	\draw[cube line] (-1,\y,\z) -- (1,\y,\z);
	\draw[cube line] (\x,\y,-1) -- (\x,\y,1);
}
\draw[y ray] (0,0,0) -- (1,-1,0) node [above right=0mm and -0.5mm]	{$\boldsymbol{y_1^-}$};
\draw[y ray] (0,0,0) -- (1,+1,0) node [left=-1.4mm]	{$\boldsymbol{y_1^+}$};
\draw[y ray] (0,0,0) -- (0,1,-1) node [below=-0.5mm]	{$\boldsymbol{y_2^-}$};
\draw[y ray] (0,0,0) -- (0,1,+1) node [above=-0.5mm]	{$\boldsymbol{y_2^+}$};
\draw[y ray] (0,0,0) -- (-1,0,1) node [above=-0.5mm]	{$\boldsymbol{y_3^-}$};
\draw[y ray] (0,0,0) -- (+1,0,1) node [above=-0.5mm]	{$\boldsymbol{y_3^+}$};

\draw[h ray] (0,0,0) -- (+1,+1,+1) node [above=-0.5mm]	{$\boldsymbol{h_0}$};
\draw[h ray] (0,0,0) -- (+1,+1,-1) node [below=-0.5mm]	{$\boldsymbol{h_1}$};
\draw[h ray] (0,0,0) -- (-1,+1,+1) node [above=-0.5mm]	{$\boldsymbol{h_2}$};
\draw[h ray] (0,0,0) -- (+1,-1,+1) node [below right=2mm and -0.5mm]	{$\boldsymbol{h_3}$};

\draw[z ray] (0,0,0) -- (0,0,1) node [above=-0.5mm]	{$\boldsymbol{z_1}$};
\draw[z ray] (0,0,0) -- (1,0,0) node [below left=-0.9mm]	{$\boldsymbol{z_2}$};
\draw[z ray] (0,0,0) -- (0,1,0) node [below right=-0.5mm]	{$\boldsymbol{z_3}$};
\end{scope}
\begin{scope}[scale=0.72, xshift=4cm, yshift=-1.6cm, text height=1.5ex,text depth=.25ex]
\def\d{1}
\node [z ray, vertex] at (0,0)					(z2)	{$\boldsymbol{z_2}$};
\node [y ray, vertex] at ($(z2)+(60:\d)$)		(y2p)	{$\boldsymbol{y_2^+}$};
\node [h ray, vertex] at ($(y2p)+(60:\d)$)		(h1)	{$\boldsymbol{h_1}$};
\node [y ray, vertex] at ($(h1)+(60:\d)$)		(y1m)	{$\boldsymbol{y_1^-}$};
\node [z ray, vertex] at ($(y1m)+(60:\d)$)		(z1)	{$\boldsymbol{z_1}$};
\node [y ray, vertex] at ($(z1)+(-60:\d)$)		(y1p)	{$\boldsymbol{y_1^+}$};
\node [h ray, vertex] at ($(y1p)+(-60:\d)$)		(h3)	{$\boldsymbol{h_3}$};
\node [y ray, vertex] at ($(h3)+(-60:\d)$)		(y3m)	{$\boldsymbol{y_3^-}$};
\node [z ray, vertex] at ($(y3m)+(-60:\d)$)		(z3)	{$\boldsymbol{z_3}$};
\node [y ray, vertex] at ($(z2)+(0:\d)$)		(y2m)	{$\boldsymbol{y_2^-}$};
\node [h ray, vertex] at ($(y2m)+(0:\d)$)		(h2)	{$\boldsymbol{h_2}$};
\node [y ray, vertex] at ($(h2)+(0:\d)$)		(y3p)	{$\boldsymbol{y_3^+}$};
\node [h ray, vertex] at ($(z1)!0.667!(h2)$)	(h0)	{$\boldsymbol{h_0}$};
\draw [to] (z2) -- (y2p) -- (h1) -- (y1m) -- (z1) -- (y1p) -- (h3) 
-- (y3m) -- (z3) -- (y3p) -- (h2) -- (y2m) -- (z2);
\draw [to] (h0) -- (y1m) -- (y1p);
\draw [to] (h0) -- (y2m) -- (y2p);
\draw [to] (h0) -- (y3m) -- (y3p);
\draw [to] (h1) to[bend right=15] (y3p);
\draw [to] (h2) to[bend right=15] (y1p);
\draw [to] (h3) to[bend right=15] (y2p);
\draw [to] (z1) to[bend right=40] (z2);
\draw [to] (z2) to[bend right=40] (z3);
\draw [to] (z3) to[bend right=40] (z1);
\end{scope}
\node [font={\small}] at (-2,1.7)	{(a)};
\node [font={\small}] at (2.9,1.7)	{(b)};
\end{tikzpicture}

%% file: fig_LevelDiagram.tex
\begin{tikzpicture}[scale=0.5,
level/.style = {very thick},
qlevel/.style = {level, red!80!black},
connect/.style = {dashed},
laser/.style = {<->, shorten <= 0.5pt, shorten >= 0.5pt, thick},
decoration = {snake, pre length=2pt, post length=6pt},
decay/.style  = {->, shorten <= 0.5pt, shorten > = 0.5pt, thick, decorate}]

\coordinate (S) at (4,0);
\coordinate (P) at (2,8.5);
\coordinate (D) at (5,5);

\draw[level] (S) -- ++(-4,0) node[left] {${}^2S_{1/2}$};
\foreach \j in {1,...,2} {
	\draw[level] ($(S)+(1,0)+(0,\j-1.5)$) -- ++(3,0) coordinate (S\j);
	\draw[connect] (S\j) ++(-3,0) -- (S); }

\draw[level] (P) -- ++(-2,0) node[left] {${}^2P_{1/2}$};
\foreach \j in {1,...,2} {
	\draw[level] ($(P)+(1,0)+(0,\j-1.5)$) -- ++(1,0) coordinate (P\j);
	\draw[connect] (P\j) ++(-1,0) -- (P); }

\draw[level] (D) -- ++(-1,0) node[left] {${}^2D_{5/2}$};
\foreach \j in {1,...,6} {
	\draw[level] ($(D)+(1,0)+0.7*(0,\j-3.5)$) -- ++(2,0) coordinate (D\j);
	\draw[connect] (D\j) ++(-2,0) -- (D); }

\node[right] at (P2) {$m_j=+1/2$};
\node[right] at (P1) {$m_j=-1/2$};

\draw[qlevel] (D3) node [right] {$\ket{2}$} -- ++(-2,0);
\draw[qlevel] (D2) node [right] {$\ket{1}$} -- ++(-2,0);
	\node[blue, right] at ($0.5*(D3)+0.5*(D2)+(1.2,0.1)$) {``dark''};
\draw[qlevel] (S1) node [right] {$\ket{0}$} -- ++(-3,0);
	\node[blue, right] at ($(S1)+(1.2,0)$) {``bright''};

\draw[laser, red!80!black] ($(S1)+(-1.3,0)$) coordinate (w1) -- ++($(D2)-(S1)$);
	\node[anchor=east, red!80!black] at ($(w1)+(0,1.6)$) {$\omega_1$};
\draw[laser, red!80!black] ($(S1)+(-0.7,0)$) coordinate (w2) -- ++($(D3)-(S1)$);
	\node[anchor=west, red!80!black] at ($(w2)+(0,1.6)$) {$\omega_2$};
\node[anchor=east, red!80!black] at ($(w1)+(-0.1,2.8)$) {\SI{729}{\nano\meter}};
\draw[laser, blue] let \p1 = ($(S)-(P)$) in ($(P)+(-1.5,0)$) -- ++(0,\y1);
\draw[decay, blue] let \p1 = ($(S)-(P)$) in ($(P)+(-1.0,0)$) -- ++(0,\y1) coordinate (w0);
	\node[anchor=west, blue] at ($(w0)+(0.2,2.3)$) {\SI{397}{\nano\meter}};
\end{tikzpicture}

%% file: fig_MeasurementScheme.tex
\begin{tikzpicture}[text height=1.5ex, text depth=.25ex
,meas/.pic={
	\draw (0,-0.2) ++ (0:0.25) arc [start angle=0, end angle=180, radius=0.25];
	\fill (0,-0.2) circle (1pt);
	\draw [-{Latex[scale=0.7]}] (0,-0.2) -- ++(70:0.45);
	\draw (0,0) +(-0.3,-0.3) rectangle +(0.3,0.3);}
,opBox2/.style={draw, thick, fill=white, minimum height=1.0cm, minimum width=0.8cm, inner sep=0pt}
,opBox3/.style={draw, thick, fill=white, minimum height=1.6cm, minimum width=0.8cm, inner sep=0pt}
,QRNGline/.style={rounded corners, double distance=1pt}
,QRNGnode/.style={draw, inner sep=2pt, rounded corners}]

\begin{scope}
\draw [very thick] 					(0.2,+0.6) 	node [left] {$\ket{1}$} -- (11.4,+0.6);
\draw [{Circle}-, very thick]		(0.2,+0.0)	node [left] {$\ket{0}$} -- (11.4,+0.0);
\draw [very thick]					(0.2,-0.6)	node [left] {$\ket{2}$} -- (11.4,-0.6);
\draw [very thick, white, dashed]	(8.6,+0.6)	-- +(1,0) ++(0,-0.6) -- +(1,0) ++(0,-0.6) -- +(1,0);

\node [opBox3] at (01.0,0)	(Uv0d)	{$U_{v_0}^\dagger$};
\node [opBox3] at (02.4,0)	(Uv1)	{$U_{v_1}$};
\node [opBox3] at (03.4,0)	(M1)	{}; \pic at (M1) {meas};
\node [opBox3] at (04.4,0)	(Uv1d)	{$U_{v_1}^\dagger$};
\node [opBox3] at (05.8,0)	(Uv2)	{$U_{v_2}$};
\node [opBox3] at (06.8,0)	(M2)	{}; \pic at (M2) {meas};
\node [opBox3] at (07.8,0)	(Uv2d)	{$U_{v_2}^\dagger$};
\node [opBox3] at (10.4,0)	(Uvl)	{$U_{v_l}$};
\node [opBox3] at (11.4,0)	(Ml)	{}; \pic at (Ml) {meas};

\draw [dashed] (-0.5,1)	node[above right] {Init.} 		rectangle ( 1.6,-1);
\draw [dashed] ( 1.8,1)	node[above right] {$A_{v_1}$}	rectangle ( 5.0,-1);
\draw [dashed] ( 5.2,1)	node[above right] {$A_{v_2}$}	rectangle ( 8.4,-1);
\draw [dashed] ( 9.8,1)	node[above right] {$A_{v_l}$}	rectangle (12.0,-1);

\def\QRNGnode#1#2#3{%
	\node [QRNGnode] at (#2,-1.7) (QRNG1) {\phantom{0 0 0 0}}
	node at (QRNG1) {\textellipsis\,#3\,\textellipsis};
	\draw [QRNGline] (QRNG1) -- ++(0,0.5) node [below left=-4pt and -1pt] {$v_#1$} -| (Uv#1);}
\QRNGnode{1}{1.0}{0 1 1 1 0 1 1 0}
\QRNGnode{2}{4.4}{1 0 0 1 0 1 1 1}
\QRNGnode{l}{9.0}{0 1 0 1 0 1 0 0}

\fill [black, opacity=0.15] (-0.5,-1.84) rectangle (10.8,-1.55);
\draw [<-] (10.9,-1.7) -- ++(0.4,0) node [right] {continuous stream from QRNG};

\end{scope}
\begin{scope}[xshift=12cm]
\draw [very thick] 	(0.8,+0.6) -- +(1.2,0) ++(0,-0.6) -- +(1.2,0) ++(0,-0.6) -- +(1.2,0);
\node [opBox3] at (1.4,0)	{$U_{v}$};
\node at (2.2,0) {$=$};

\draw [very thick] 	(2.4,+0.6) -- +(2.2,0) ++(0,-0.6) -- +(2.2,0) ++(0,-0.6) -- +(2.2,0);
\node [opBox2] at (3,+0.3)	{$R^{(1)}_v$};
\node [opBox2] at (4,-0.3)	{$R^{(2)}_v$};
\end{scope}
\end{tikzpicture}

%% file: fig_CorrelatorResults.tex
% !TeX root = testFigs.tex

\pgfplotsset{CorrResults/.style={
	width=23mm, y=3.5mm, scale only axis, clip=false,
	xtick pos=left, ytick pos=left, ytick=data,
	visualization depends on={value \thisrow{label} \as \labela},
	mark size=1pt, nodes near coords align={horizontal},
%	node near coords style={left=-2pt},
	xticklabel style={/pgf/number format/fixed,
		/pgf/number format/fixed zerofill,
		/pgf/number format/precision=2}}}

\begin{tikzpicture}[every node/.style={font=\footnotesize}]
%\draw  (-6.15,-0.4) rectangle ++(8.6,9.5);
\begin{axis}[CorrResults, name=exp2,
	xmin=-0.34, xmax=-0.292, xtick distance=0.02, ymin=-24.8,	ymax=-0.4,
	yticklabels={$\expval*{A_{y_1^-}A_{y_1^+}}$, $\expval*{A_{y_1^-}A_{h_1}}$, $\expval*{A_{y_1^-}A_{h_0}}$, 
		$\expval*{A_{y_1^-}A_{z_1}}$, $\expval*{A_{y_2^-}A_{y_2^+}}$, $\expval*{A_{y_2^-}A_{h_2}}$,
		$\expval*{A_{y_2^-}A_{h_0}}$, $\expval*{A_{y_2^-}A_{z_2}}$, $\expval*{A_{y_3^-}A_{y_3^+}}$,
		$\expval*{A_{y_3^-}A_{h_3}}$, $\expval*{A_{y_3^-}A_{h_0}}$, $\expval*{A_{y_3^-}A_{z_3}}$,
		$\expval*{A_{y_1^+}A_{h_2}}$, $\expval*{A_{y_1^+}A_{h_3}}$, $\expval*{A_{y_1^+}A_{z_1}}$,
		$\expval*{A_{y_2^+}A_{h_1}}$, $\expval*{A_{y_2^+}A_{h_3}}$, $\expval*{A_{y_2^+}A_{z_2}}$,
		$\expval*{A_{y_3^+}A_{h_1}}$, $\expval*{A_{y_3^+}A_{h_2}}$, $\expval*{A_{y_3^+}A_{z_3}}$,
		$\expval*{A_{z_1}A_{z_2}}$, $\expval*{A_{z_1}A_{z_3}}$, $\expval*{A_{z_2}A_{z_3}}$}]
\draw[dashed] (-0.3333,\pgfkeysvalueof{/pgfplots/ymin})
	-- (-0.3333,\pgfkeysvalueof{/pgfplots/ymax});
\draw[Stealth-] (-0.3333,\pgfkeysvalueof{/pgfplots/ymax}) |- ++(4mm,3mm)
	node[anchor=west] {\num{-0.333}};
\addplot+[only marks, error bars/.cd, x dir=both, x explicit] 
	table[x=x,y=y,x error=error] {
		x			error	y	label
		-0.329490	0.0012	-1	\num{-0.329(2)}
		-0.329641	0.0012	-2	\num{-0.330(2)}
		-0.324904	0.0012	-3	\num{-0.325(2)}
		-0.329401	0.0012	-4	\num{-0.329(2)}
		-0.325738	0.0012	-5	\num{-0.326(2)}
		-0.329330	0.0012	-6	\num{-0.329(2)}
		-0.331460	0.0012	-7	\num{-0.331(2)}
		-0.331510	0.0012	-8	\num{-0.332(2)}
		-0.326387	0.0012	-9	\num{-0.326(2)}
		-0.328821	0.0012	-10	\num{-0.329(2)}
		-0.329561	0.0012	-11	\num{-0.330(2)}
		-0.331541	0.0012	-12	\num{-0.332(2)}
		-0.331451	0.0012	-13	\num{-0.331(2)}
		-0.332154	0.0012	-14	\num{-0.332(2)}
		-0.330097	0.0012	-15	\num{-0.330(2)}
		-0.326537	0.0012	-16	\num{-0.327(2)}
		-0.326913	0.0012	-17	\num{-0.327(2)}
		-0.329001	0.0012	-18	\num{-0.329(2)}
		-0.327255	0.0012	-19	\num{-0.327(2)}
		-0.328988	0.0012	-20	\num{-0.329(2)}
		-0.329319	0.0012	-21	\num{-0.329(2)}
		-0.328698	0.0012	-22	\num{-0.329(2)}
		-0.328380	0.0012	-23	\num{-0.328(2)}
		-0.332498	0.0012	-24	\num{-0.332(2)}
	};
\addplot [nodes near coords={\labela}, only marks, mark=none] 
	table[x expr=\pgfkeysvalueof{/pgfplots/xmax}, y=y] {
		x			error	y	label
		-0.329490	0.0012	-1	\num{-0.329(2)}
		-0.329641	0.0012	-2	\num{-0.330(2)}
		-0.324904	0.0012	-3	\num{-0.325(2)}
		-0.329401	0.0012	-4	\num{-0.329(2)}
		-0.325738	0.0012	-5	\num{-0.326(2)}
		-0.329330	0.0012	-6	\num{-0.329(2)}
		-0.331460	0.0012	-7	\num{-0.331(2)}
		-0.331510	0.0012	-8	\num{-0.332(2)}
		-0.326387	0.0012	-9	\num{-0.326(2)}
		-0.328821	0.0012	-10	\num{-0.329(2)}
		-0.329561	0.0012	-11	\num{-0.330(2)}
		-0.331541	0.0012	-12	\num{-0.332(2)}
		-0.331451	0.0012	-13	\num{-0.331(2)}
		-0.332154	0.0012	-14	\num{-0.332(2)}
		-0.330097	0.0012	-15	\num{-0.330(2)}
		-0.326537	0.0012	-16	\num{-0.327(2)}
		-0.326913	0.0012	-17	\num{-0.327(2)}
		-0.329001	0.0012	-18	\num{-0.329(2)}
		-0.327255	0.0012	-19	\num{-0.327(2)}
		-0.328988	0.0012	-20	\num{-0.329(2)}
		-0.329319	0.0012	-21	\num{-0.329(2)}
		-0.328698	0.0012	-22	\num{-0.329(2)}
		-0.328380	0.0012	-23	\num{-0.328(2)}
		-0.332498	0.0012	-24	\num{-0.332(2)}
	};
\end{axis}

\begin{axis}[CorrResults, name=exp1, 
	at={(exp2.north west)}, xshift=-18mm, anchor=north east,
	xmin=0.3295, xmax=0.343,  ymin=-13.8,	ymax=-0.4,
	yticklabels={$\expval*{A_{y_1^-}}$, $\expval*{A_{y_2^-}}$, $\expval*{A_{y_3^-}}$, 
		$\expval*{A_{y_1^+}}$, $\expval*{A_{y_2^+}}$, $\expval*{A_{y_3^+}}$, $\expval*{A_{h_1}}$,
		$\expval*{A_{h_2}}$, $\expval*{A_{h_3}}$, $\expval*{A_{h_0}}$, $\expval*{A_{z_1}}$,
		$\expval*{A_{z_2}}$, $\expval*{A_{z_3}}$},
	xticklabel style={/pgf/number format/fixed, /pgf/number format/fixed zerofill,
		/pgf/number format/precision=3}]
\draw[dashed] (0.3333,\pgfkeysvalueof{/pgfplots/ymin}) -- (0.3333,\pgfkeysvalueof{/pgfplots/ymax});
\draw[Stealth-] (0.3333,\pgfkeysvalueof{/pgfplots/ymax}) |- ++(4mm,3mm) node[anchor=west] {0.3333};
\addplot+[only marks, error bars/.cd, x dir=both, x explicit] 
	table[x=x,y=y,x error=error] {
		x			error	y	label
		0.332379	0.00046	-1	\num{0.3324(5)}
		0.332841	0.00046	-2	\num{0.3328(5)}
		0.331979	0.00046	-3	\num{0.3320(5)}
		0.333071	0.00046	-4	\num{0.3331(5)}
		0.334294	0.00046	-5	\num{0.3343(5)}
		0.334174	0.00046	-6	\num{0.3342(5)}
		0.332070	0.00046	-7	\num{0.3321(5)}
		0.332229	0.00046	-8	\num{0.3322(5)}
		0.333174	0.00046	-9	\num{0.3332(5)}
		0.333504	0.00046	-10	\num{0.3335(5)}
		0.334014	0.00046	-11	\num{0.3340(5)}
		0.332718	0.00046	-12	\num{0.3327(5)}
		0.333041	0.00046	-13	\num{0.3330(5)}
	};
\addplot [nodes near coords={\labela}, only marks, mark=none] 
	table[x expr=\pgfkeysvalueof{/pgfplots/xmax}, y=y] {
		x			error	y	label
		0.332379	0.00046	-1	\num{0.3324(5)}
		0.332841	0.00046	-2	\num{0.3328(5)}
		0.331979	0.00046	-3	\num{0.3320(5)}
		0.333071	0.00046	-4	\num{0.3331(5)}
		0.334294	0.00046	-5	\num{0.3343(5)}
		0.334174	0.00046	-6	\num{0.3342(5)}
		0.332070	0.00046	-7	\num{0.3321(5)}
		0.332229	0.00046	-8	\num{0.3322(5)}
		0.333174	0.00046	-9	\num{0.3332(5)}
		0.333504	0.00046	-10	\num{0.3335(5)}
		0.334014	0.00046	-11	\num{0.3340(5)}
		0.332718	0.00046	-12	\num{0.3327(5)}
		0.333041	0.00046	-13	\num{0.3330(5)}
	};
\end{axis}

\begin{axis}[CorrResults, name=exp3,
	at={(exp2.south west)}, xshift=-18mm, anchor=south east,
	xmin=-1.02, xmax=-0.88, ymin=-3.8,	ymax=-0.4, 
	yticklabels={$\expval*{A_{z_1}A_{y_1^-}A_{y_1^+}}$, $\expval*{A_{z_2}A_{y_2^-}A_{y_2^+}}$,
		$\expval*{A_{z_3}A_{y_3^-}A_{y_3^+}}$}]
\draw[dashed] (-1,\pgfkeysvalueof{/pgfplots/ymin})
	-- (-1,\pgfkeysvalueof{/pgfplots/ymax});
\addplot+[only marks, error bars/.cd, x dir=both, x explicit] 
	table[x=x,y=y,x error=error] {
		x	error	y	label
		-0.984856	0.000454	-1	\num{-0.9849(5)}
		-0.984185	0.000463	-2	\num{-0.9842(5)}
		-0.984493	0.000458	-3	\num{-0.9845(5)}
	};
\addplot [nodes near coords={\labela}, only marks, mark=none] 
	table[x expr=\pgfkeysvalueof{/pgfplots/xmax}, y=y] {
		x	error	y	label
		-0.984856	0.000454	-1	\num{-0.9849(5)}
		-0.984185	0.000463	-2	\num{-0.9842(5)}
		-0.984493	0.000458	-3	\num{-0.9845(5)}
	};
\end{axis}
\end{tikzpicture}

%% file: main.bbl

%% file: supp_content.tex
% !TeX root = main.tex
% !TeX spellcheck = en_US

\iftoggle{arXiv}{
	\clearpage
	{\LARGE\bfseries Supplementary Material}
}{}

\section{Glossary}
In the following sections we will use terminology which can have different connotations in different contexts. We give our definitions here for clarity:
\begin{description}
\item[Projection noise] Also known as quantum projection noise or shot noise, this refers to the stochastic uncertainty inherent to finite numbers of measurements in quantum mechanics \cite{93Itano}. For dichotomic observables this is well described by a Bernoulli distribution of mean $\bar{p}\in[0,1]$ and variance $\bar{p}(1-\bar{p})$. The uncertainty of the mean for a measurement in which projection noise is the only source of noise is given by $\sqrt{\bar{p}(1-\bar{p})/N}$, where $N$ is the number of measurements on the observable.
\item[Repeatability] This is characterized by the probability for obtaining different outcomes in two consecutive measurements of an observable $A_u$, which is defined in Eq.~(\ref{eq:rep1}).
%TODO Add description for extended repeatability (with different name).
\item[Sharpness] Sharp measurements are minimally disturbing \iftoggle{bibrun}{}{\cite{14Chiribella}} and their outcomes are the same if performed repeatedly. Note that sharpness implies that measurements are repeatable even when other sharp compatible measurements are performed in between two successive realizations of the same measurement. In quantum theory, sharp measurements are represented by self-adjoint operators; ``ideal measurements'' as defined by von Neumann \iftoggle{bibrun}{}{\cite{32vonNeumann}} are sharp measurements.
\item[Compatibility] Generally, two observables are compatible if measurements on them yield the same result regardless of whether they are performed simultaneously or sequentially in any temporal order, and regardless of the number of times they are measured. In QM, two sharp measurements are compatible if their operators commute. In the orthogonality graph for the Yu-Oh set in FIG.~\iftoggle{arXiv}{\ref{fig:YuOhRays}}{\ref{m-fig:YuOhRays}}b, the 24 edges connect the 24 supposedly compatible pairs of measurements. Note that perfect compatibility is impossible to achieve in real experiments, unless the locality and detection loopholes are simultaneously closed.
\item[Context signaling] Context signaling is present in a dataset if the statistics of an observable are significantly different depending on the context in which it is measured, i.e.  what compatible observable it is measured together with. The existence of context signaling precludes compatibility.
\item[Context signaling backwards in time] This is the particular manifestation of context signaling when the context observable is measured after the main observable in a sequential measurement. It is characterized by Eq.~(\ref{eq:Signaling}a) and should be zero down to statistical uncertainties under the assumption of causality. It is a useful reference to quantify the presence of context signaling forwards in time \cite{16Cabello2}.
\item[Context signaling forwards in time] This is the particular manifestation of context signaling when the context observable is measured before the main observable in a sequential measurement. It is characterized by Eq.~(\ref{eq:Signaling}b) and should be zero down to statistical uncertainties if observable and contexts are compatible. 
\end{description}

\section{Qutrit coherence}
\begin{figure}
\includegraphics[width= 0.95\columnwidth]{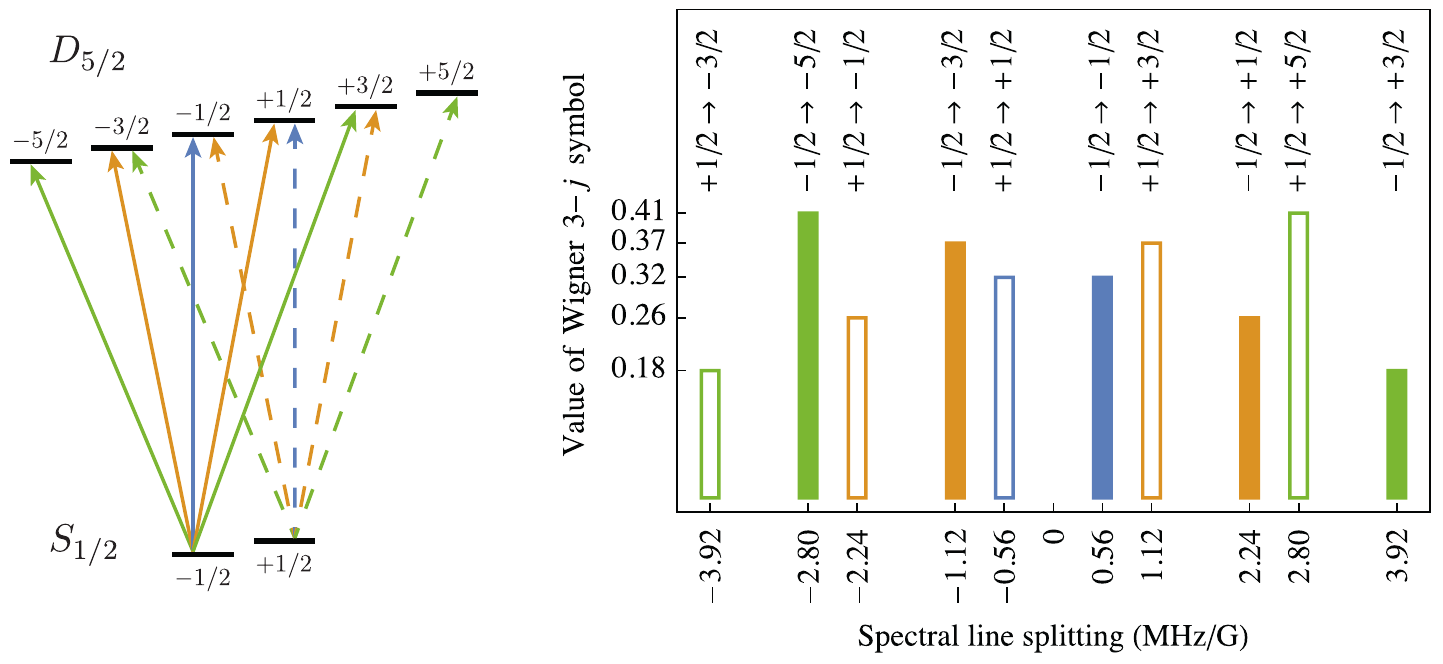}
\caption{(color online) Spectral lines on the $S_{1/2}\leftrightarrow D_{5/2}$ quadrupole transition at \SI{729}{nm} for a $^{40}$Ca$^+$ ion in a magnetic field (blue/orange/green arrows and bars for $|\Delta m| = 0/1/2$). The frequencies are Zeeman shifted by the given values in MHz/G. The Rabi frequencies of the lines are proportional to the Wigner 3-$j$ symbol \cite{ThLeupold}. The full/empty bars on the right correspond to the solid/dashed arrows on the left.}
\label{fig:Transitions}
\end{figure}

In our experiments, qutrit states $\ket{1}$ and $\ket{2}$ are encoded into Zeeman sublevels of the metastable $D_{5/2}$ manifold of the $^{40}$Ca$^+$ ion, with a lifetime of $\approx\SI{1}{s}$, whereas $\ket{0}$ is one of the $S_{1/2}$ ground states. The coherence between these states is determined by the stability of the  magnetic field defining the quantization axis of our system, and of the \SI{729}{nm} laser with which we address the $\ket{0}\leftrightarrow\ket{1}$ and $\ket{0}\leftrightarrow\ket{2}$ transitions. Applying Ramsey techniques, we measure coherence times of $\approx\SI{12}{ms}$ for the $\ket{1}\leftrightarrow\ket{2}$ transition and $\approx\SI{2.5}{ms}$ for the
$\ket{0}\leftrightarrow\ket{1}$ and $\ket{0}\leftrightarrow\ket{2}$ transitions. This indicates that our main source of noise leads to fluctuations which are common mode to levels $\ket{1}$ and $\ket{2}$. Since the frequencies of the $\ket{0}\leftrightarrow\ket{1}$ and $\ket{0}\leftrightarrow\ket{2}$ transitions have opposite dependencies on the magnetic field (FIG.~\ref{fig:Transitions}b), we suspect that laser-frequency fluctuations are the largest contributor to the observed loss of coherence.

\section{Experimental details}\label{sec:expdet}
The sustained sequence demonstrated in this work represents an experimental paradigm which motivated some changes to technologies and techniques typically used in trapped-ion experiments. The most profound technological change was an upgrade of the control system \cite{17Bermudez} to increase the maximum number of measurements in a subsequence from under \num{60} to over \num{1000} and to integrate the quantum random number generator.

Additionally, we work with detection settings that make dedicated Doppler-cooling pulses unnecessary. Fluorescence state-detection and Doppler cooling both make use of a $<\SI{1}{MHz}$-linewidth laser at \SI{397}{nm}, near-resonant with the $S_{1/2}\leftrightarrow P_{1/2}$ transition (natural linewidth $\Gamma\approx (2\pi)\,\SI{21}{MHz}$) \cite{ThLeupold}. Detection is normally performed with the laser tuned close to resonance, while Doppler cooling is optimal when the laser is detuned by half a transition linewidth. For this work, we red-detune the detection laser by half a linewidth from resonance to ensure that the detection pulse cools the motion close to the Doppler limit \cite{03Leibfried2} whenever there is a bright detection. This comes at a cost in the number of photons collected during bright detection in a window of fixed length, but removes the need for dedicated Doppler-cooling pulses, thereby shortening the overall experiment duration. Working at the Doppler temperature rather than in the motional ground-state has an impact on the quality of coherent rotations. The infidelity of $\pi$-pulses on the \SI{729}{nm} transitions is thereby bounded from below by $\num{1e-3}$ even under otherwise ideal conditions. A finer estimate for the pulse infidelity can be obtained from the probabilities
\begin{align}\label{eq:PInf}
P_{v;z_1}^\text{(inf)}=\frac{N_{z_1}(A_v{=}{-}1)}{N_{z_1}(A_v)},
\end{align}
where $N_{z_1}(A_v{=}{-}1)\equiv N(A_{z_1}{=}{-}1,A_v{=}{-}1)$, for detecting a ``bright'' qutrit state along $v=z_2$ and $v=z_3$ after the qutrit was in $\ket{0}$ (along $z_1$). Ideally, these probabilities would be zero; instead we find them to be $\approx\num{5e-3}$, which gives the infidelity of $\pi$-pulses on the $\ket{0}\leftrightarrow\ket{1}$ and $\ket{0}\leftrightarrow\ket{2}$ transitions. When we analyze all supposedly compatible pairs we find an average value $\num{37(11)e-4}$, and the largest deviation from zero is $P_{y_2^-;y_2^+}^\text{(inf)}\approx\num{59e-4}$.

Here and for later use, we define numbers of occurrence in the dataset conditioned on the preceding projection onto a specific state $i\in V$ as $N_i(A_v{=}a_1)\equiv N(A_i{=}{-}1,A_v{=}a_1)$ etc. Conversely, $N_{\perp i}(A_v{=}a_1)\equiv N(A_i{=}{+}1,A_v{=}a_1)$ etc. are numbers of occurrence following the projection of the state into the plane normal to $i$. Furthermore, omitting an outcome in the arguments list of $N$ means summing over both outcomes, i.e. $N(A_v)\equiv\sum_{a_1} N(A_v{=}a_1)$.

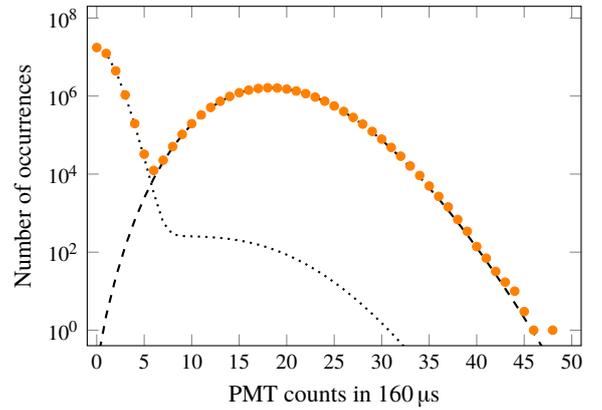
\begin{figure}
\input{fig_DetectionHistogram}
\caption{Histogram of the PMT counts of all detection events in the experimental sequence (orange dots). The black lines represent scaled Poisson distributions, whose sum was fitted to the experimental data. The 53 million detection events are split between \num{3.56e7} occurrences in the Poisson distribution with an average of \SI{0.709}{cts} (``dark''), and \num{1.78e7} in the Poisson distribution with an average of \SI{18.75}{cts} (``bright'').}
\label{fig:Detection}
\end{figure}

FIG.~\ref{fig:Detection} shows a histogram of the Photo-Multiplier-Tube (PMT) outputs and includes all 53 million detection events. We fit the histogram using a simplified model consisting of two Poisson distributions. We infer the state by setting a threshold at \SI{5.5}{counts}, close to the crossing point of the distributions. From a detection event in which the number of PMT counts is higher (lower) than this threshold, we infer the state to be bright (dark). From simulations of decay of the dark state and relevant scattering rates and detection time we estimate an error of \num{1.9e-4} for both identifying a $D_{5/2}$-state as ``bright'' and a $S_{1/2}$-state ($\ket{0}$) as ``dark''; the contribution from the $D_{5/2}$ spontaneous decay during a dark detection is \num{1e-4} \cite{08Myerson}. Another quantity of interest is the ``detection repeatability'', given by
\begin{subequations}
\begin{align}
P\pqty\big{\text{``dark''}  |S_{1/2}}
	&\approx\frac{N_{z_1}(A_{z_1}{=}{+}1)}{N_{z_1}(A_{z_1})} \\
P\pqty\big{\text{``bright''}|D_{5/2}}
	&\approx\frac{N_{\perp z_1}(A_{z_1}{=}{-}1)}{N_{\perp z_1}(A_{z_1})},
\end{align}
\end{subequations}
which both give $\approx\num{5e-4}$. Compared to infidelities of coherent operations, detection errors are thus of minor influence.

\section{Sharpness and Compatibility}

In this section we characterize the extent to which the underlying assumptions of the SIC test are fulfilled. These are measurement sharpness and compatibility.

Sharp measurements are repeatable and minimally disturbing. To quantify how well this is realized experimentally, we calculate the probability
\begin{equation}
\label{eq:rep1}
R_{u}=\frac{\sum_{a_1} N(A_u{=}a_1,A_u{=}a_1)}{N(A_u,A_u)}
\end{equation}
for obtaining the same outcomes in two consecutive measurements of observable $A_u$. Ideally, this probability would be 1 for all observables. Instead, we experimentally find them to be lower with deviations on the order of \num{e-3}, and $R_{h_1}=\num{0.9967(2)}$ being the lowest. 

\begin{figure}
\input{fig_Repeatability}
\caption{Measure of the repeatability of our measurements calculated according to Eq.~(\ref{eq:rep1}). The  ideal value is 1.}
\label{fig:Repeatability}
\end{figure}
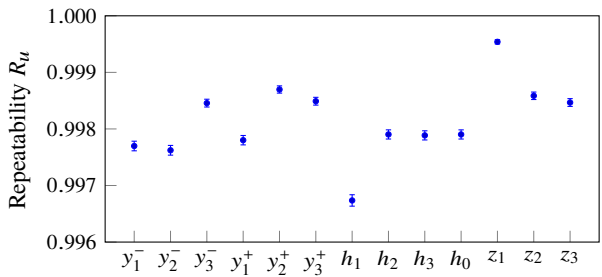

\begin{figure}
\input{fig_Signaling}
\caption{Measures for context signaling backwards (sba, orange) and forwards (sfo, blue) in time. The histograms represent the distributions of the probabilities calculated according to Eqs.~\eqref{eq:Signaling}. Normal distributions are fit to them to find their centers $\mu^\text{(ba)}=\num{-0.02(4)}$ and $\mu^\text{(fo)}=\num{-0.04(4)}$, and standard deviations $\sigma^\text{(ba)}=\num{0.93(4)}$ and $\sigma^\text{(fo)}=\num{1.64(4)}$. The results are consistent with no context signaling backward in time and projection noise the only source of noise present. The increase in $\sigma^\text{(fo)}$ with respect to $\sigma^\text{(ba)}$ is consistent with systematic errors in coherent operations. These lead to input-state-dependent context signaling forwards in time. The uncertainties assigned to fitting parameters are estimated based on non-weighted least-squares regression.}
\label{fig:Signaling}
\end{figure}
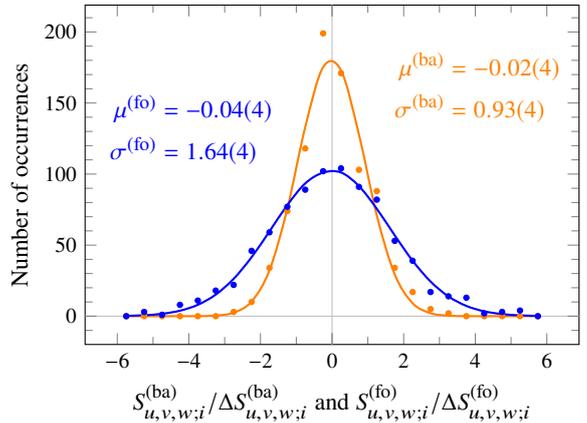

Next, we examine the compatibility of our measurements, for which a necessary condition is the absence of context signaling. To quantify context signaling in the data we define
\begin{subequations}	\label{eq:Signaling}
	\begin{align}
	S_{u,v,w;i}^\text{(ba)}&=\frac{N_i(A_u{=}{-}1,A_v)}{N_i(A_u,A_v)}
	-\frac{N_i(A_u{=}{-}1,A_w)}{N_i(A_u,A_w)},	\\
	S_{u,v,w;i}^\text{(fo)}&=\frac{N_i(A_v,A_u{=}{-}1)}{N_i(A_v,A_u)}
	-\frac{N_i(A_w,A_u{=}{-}1)}{N_i(A_w,A_u)}
	\end{align}
\end{subequations}
for signaling backwards (ba) and forwards (fo) in time. Note that these are defined for definite input states $i$ because for mixed input states, there is no difference between compatible and non-compatible preceding measurements. Looking at signaling backwards in time first, values of $S_{u,v,w;i}^\text{(ba)}$ significantly different from zero would imply that the probability of measuring $A_u$ to be dark is affected by the following (ideally) compatible measurement being along $A_v$ rather than $A_w$. This corresponds to context signaling backwards in time and should not occur under the assumption of causality \cite{16Cabello2}. Figure~\ref{fig:Signaling} shows the histogram of $S_{u,v,w;i}^\text{(ba)}/\Delta S_{u,v,w;i}^\text{(ba)}$ for all combinations of rays where $v$ and $w\neq v$ are compatible with $u$. The standard errors  $\Delta S_{u,v,w;i}^\text{(ba)}$ of the $S_{u,v,w;i}^\text{(ba)}$ were calculated from projection noise assuming a Bernoulli distribution \cite{93Itano}. The distribution is consistent with normally distributed values centered at zero and with unit standard deviation. This means that our data are consistent with no context signaling backwards in time and that the variances we observe can be explained by projection noise.

Proceeding analogously for context signaling forwards in time, we find that the histogram values for $S_{u,v,w;i}^\text{(fo)}/\Delta S_{u,v,w;i}^\text{(fo)}$ are fitted by a normal distribution centered at $\mu^\text{(fo)}=\num{-0.04(4)}$ with a standard deviation $\sigma^\text{(fo)}=\num{1.64(4)}$. For determining the uncertainties $\Delta S_{u,v,w;i}^\text{(fo)}$ we assumed that there is no correlation between the errors in both terms in Eq.~(\ref{eq:Signaling}b). This is a realistic assumption if projection noise is the only source of uncertainty, but not necessarily the case in the presence of systematic coherent under- or over-rotations. The fact that the mean is consistent with zero does not necessarily grant that there is no significant context signaling forwards in time. To verify this, we perform this same study for each input state independently and find identical distributions down to statistical uncertainties. We have also investigated the subset of points lying at the wings of the distribution, and found no evidence that specific values of $\{u,v,w;i\}$ contribute prominently to their composition.

Together, these measures quantify to which degree our implementations of observables in the Yu-Oh set are sharp and compatible. Experimental imperfections unavoidably lead to deviations from the ideal values. We believe that our small but finite non-repeatability and context signaling should reflect on the contextuality witnesses in Eqs.~(\iftoggle{arXiv}{\ref{eq:SICWitnesses}}{\ref{m-eq:SICWitnesses}}) but are, as stated, not aware of a suitable penalizing algorithm.

\section{Memory needed for simulating a sequential Yu-Oh experiment}

\begin{figure}
\includegraphics[width= 0.95\columnwidth]{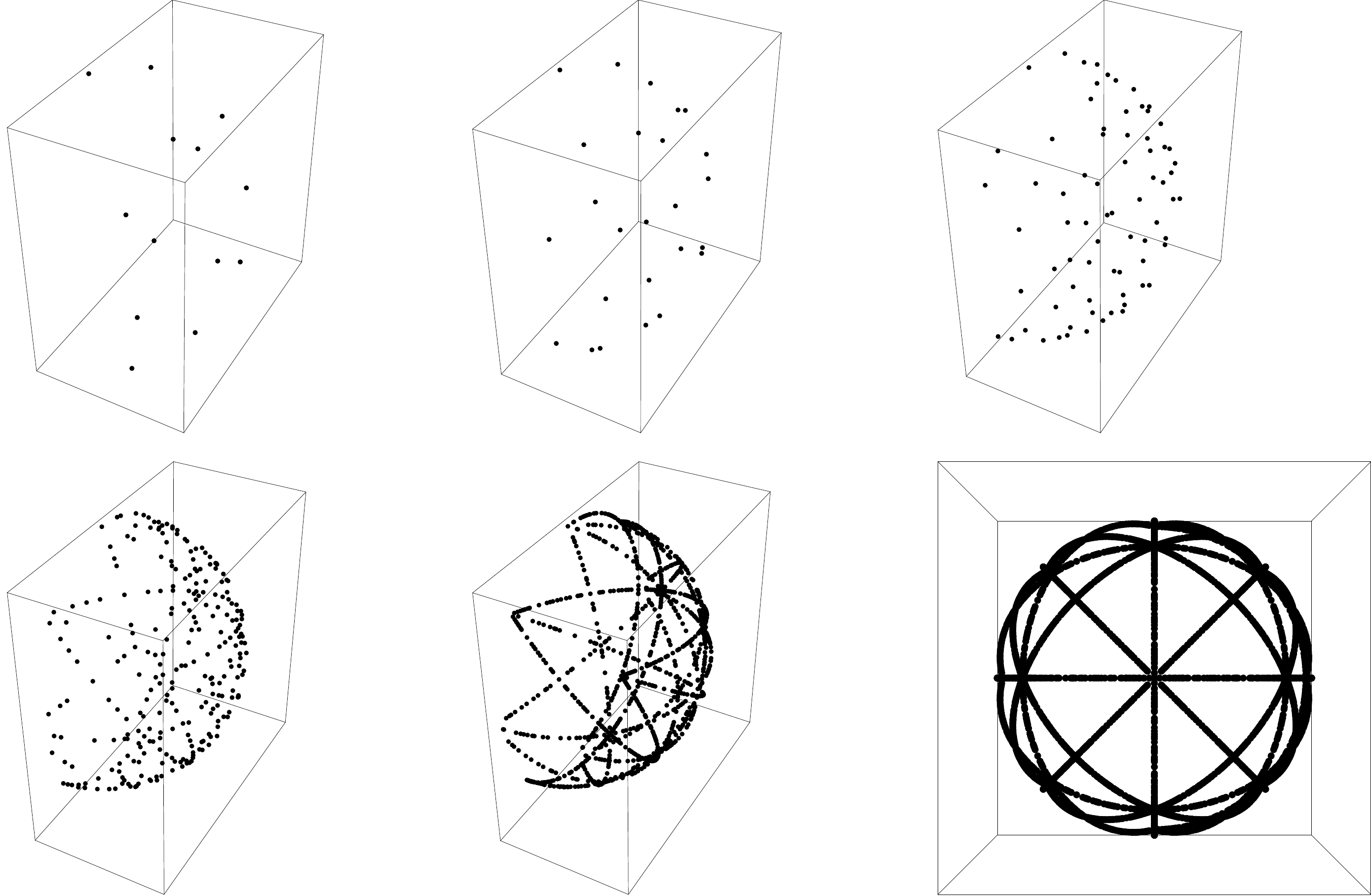}
\caption{Assuming that we start counting when the qutrit in one of the 13 quantum states of the Yu-Oh set (represented by the dots over a semisphere in the upper left), the successive figures, from left to right and from up to down, show the possible post-measurement quantum states after one, two, three, four, and five measurements, respectively. The number of possible post-measurement states is 25, 73, 265, 1033, and 3649, respectively, and always increases with the number of time steps. Notice that all the states lie into one of the 13 semicircles corresponding to the states with real components orthogonal to the 13 states of the Yu-Oh set.}
\label{fig:Memory}
\end{figure}

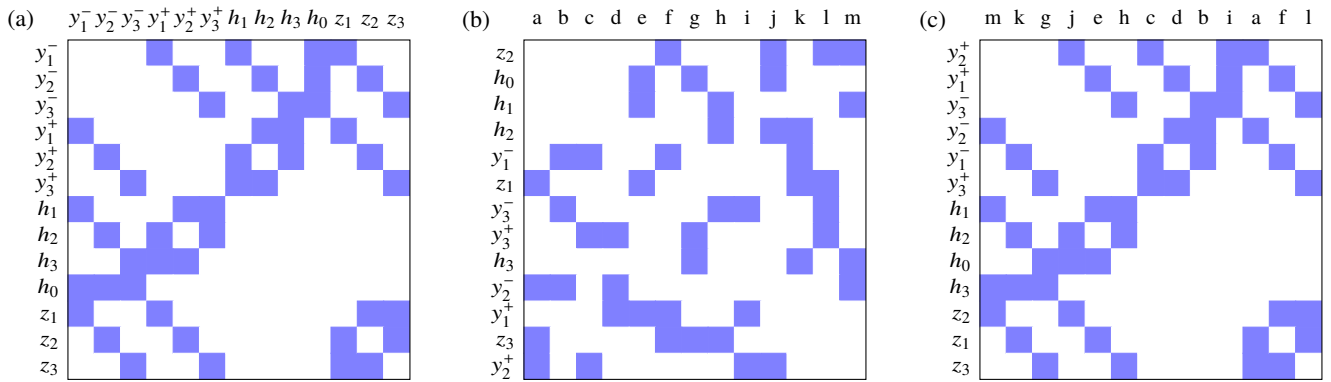
\begin{figure*}
\input{fig_Blind}
\caption{Correlation matrices showing the compatibility structure of the observables in the Yu-Oh set. Blue squares mark compatible pairs. (a) Ideal correlation matrix using the labeling convention for vectors $v$ defined in TABLE~\iftoggle{arXiv}{\ref{tab:RayAngles}}{\ref{m-tab:RayAngles}}. (b) Initial matrix with randomized ray identifiers $a$ to $m$ assigned to the ``true'' (but for the blind analysis ignored) vector labels (left). (c) Applying the re-ordering steps described in the text, one obtains a compatibility structure equivalent to the one in (a), but possibly different order of vectors. The compatibility graphs for (a) and (c) are hence the same.}
\label{fig:BlindDetermination}
\end{figure*}

Simulating quantum contextuality with classical systems requires memory \cite{11Kleinmann}. It has been recently proven \cite{17Cabello} that the minimum number of bits of memory a classical machine must have to simulate the predictions of QM for an ideal experiment with sequential measurements randomly chosen from a SIC set of measurements is given by
\begin{equation}\label{eq:BitsMemory}
-\sum_i p_i \log_2 p_i, 
\end{equation}
where $p_i$ is the probability of occurrence of each of the quantum states achievable during the experiment.

For the case that the SIC set is the Yu-Oh set, and assuming that we start counting states when the qutrit is in one of the 13 quantum states of the Yu-Oh set, the number of possible quantum states after one, two, three, four, and five measurements is 25, 73, 265, 1033, and 3649, respectively (FIG.~\ref{fig:Memory}). However, not all states are equally likely to be occupied. If we take into account their probabilities of occurrence, Eq.~(\ref{eq:BitsMemory}) implies that the memory a classical system would need is significantly higher than the classical information carrying capacity a qutrit has. For example, considering just four measurements (for which analytical expressions exist), we can already say that the memory a classical system requires to simulate the quantum predictions has to be higher than $5.529$ bits. From this point of view the choice of the Yu-Oh versus, e.g., the SIC set of Peres and Mermin \cite{90Peres,90Mermin} is justified: not only does the Yu-Oh set contain qutrit measurements (rather than two-qubit measurements), but also the classical simulation of the quantum predictions requires more memory (the Peres-Mermin set requires $\log_2 24 \approx 4.585$ bits of memory) \cite{17Cabello}.

\section{Blind determination of compatibility structure}

The full compatibility structure of the Yu-Oh SIC set (FIG.~\iftoggle{arXiv}{\ref{fig:YuOhRays}}{\ref{m-fig:YuOhRays}}b) can be inferred from a clean enough dataset \emph{without invoking QM}. For $\pqty{A_u,A_w}$ to be compatible, $\epsilon_{u,w;i}$ must be equal to 0 for all input states $\ket{i}$. Discriminating experimental results for $\epsilon_{u,w;i}$, say at \num{0.05}, allows one to find all combinations of compatible observables, i.e.  edges in the compatibility graph.

We demonstrate a full reconstruction of the compatibility graph starting from a random assignment of ray identifiers $a$ to $m$ to experimentally measured observables, and blindly analyzing the data purely based on the compatibility properties of the observables. Applying the discrimination described above, one finds a compatibility matrix (FIG.~\ref{fig:BlindDetermination}b) that does not resemble the one for the ordered case. The task now is to recover the latter, which can be done in four steps:
\begin{enumerate}
\item If a row only shows three blue squares (corresponding to edges in FIG.~\iftoggle{arXiv}{\ref{fig:YuOhRays}}{\ref{m-fig:YuOhRays}}b), it belongs to an $h_\alpha$ ray. All these rows are put into the block of $h_\alpha$ rays as in FIG.~\ref{fig:BlindDetermination}a, despite their order being unknown.
\item If a row has an edge with an $h_\alpha$ ray, it belongs to a $y_k^\sigma$ ray, and can be sorted to the left (or above) the block of $h_\alpha$ rays.
\item The $z_k$ ray are now automatically at the end of the list, as in FIG.~\ref{fig:BlindDetermination}a. They cannot be distinguished any further, which reflects the symmetry of the compatibility graph. We fix their order as is and sort the $y_k^\sigma$ rays accordingly, as only the $y_1^\sigma$ rays have an edge with $z_1$, etc.
\item The $h_\alpha$ cannot be distinguished any further either. But we can define the rightmost (bottommost) ray to be the one that has edges to all $y_k^-$ as does the original $h_0$ ray. This allows moving all $y_k^-$ to the left (top) of the list. The remaining $h_k$ rays can then be ordered such that the first $y_k^-$ has an edge with the first $h_k$ in the list etc.
\end{enumerate}

Performing all these steps, one always reaches a compatibility matrix and equivalently graph that equals the ``true'' matrix and graph, albeit the indexes possibly being permuted. This highlights the fact that a given set of rays determines uniquely the compatibility graph, but not vice versa \iftoggle{bibrun}{}{\cite{12Yu}}.

\section{``Standard'' violations of SIC}

\begin{figure}
\input{fig_YOSh}
\caption{Experimental results for SIC witnesses $\expval*{\chi_\text{YO}}_i$ and $\expval*{\chi_\text{opt3}}_i$ for specific input states along rays $i\in V$ (see text for details on their calculation). Error bars reflect shot noise; dashed lines represent values predicted by quantum mechanics; the left-hand values of the diagram's plot ranges are the bounds for non-contextual correlations.}
\label{fig:YOSh}
\end{figure}
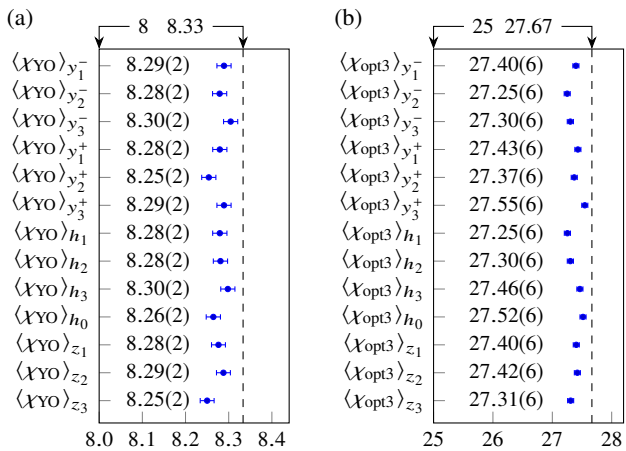
Our experiment sequence was designed to violate SIC inequalities by  performing a single long series of measurements of randomly chosen observables by recycling the states from previous measurements. However, the dataset obtained through this procedure can be evaluated in a way that corresponds to the ``standard approach'' to experimental demonstrations of the violation of SIC inequalities where measurements are repeatedly performed on specifically prepared states of the system.

For this purpose, we define averages $\expval*{\dots}_i$ in full analogy to all previous definitions $\expval*{\dots}$, but based on the numbers of occurrence $N_i$ conditioned on a preceding projection onto the ray $i$ (see Section~\ref{sec:expdet}). The results of the SIC witnesses $\expval*{\chi_\text{YO}}_i$ and $\expval*{\chi_\text{opt3}}_i$ for input states along all rays $i\in V$ approach the QM limits and are incompatible with the non-contextual bounds by multiple standard deviations, at least \num{15} and \num{43}, respectively (FIG.~\ref{fig:YOSh}).

\section{Data repository}

The complete raw dataset is publicly available from an open repository on \url{http://www.tiqi.ethz.ch/publications-and-awards/public-datasets.html}. We encourage readers who want to expand our work with further data analysis or representations to do so. As described in the main part of the paper, the dataset does not include the data purged in real time by the computer system due to long sequences of dark measurements. 

While the full dataset contains 54.5 million measurements, we had to ignore about 1 million measurements, yielding the almost 53.5 million measurements evaluated in the paper. The reason for this was an error in the live calculation of the ideal detection threshold during the experimental sequence. The computer control system consistently thresholded at 4.5 counts, whereas the ideal value was 5.5 counts (Sec.~\ref{sec:expdet}). Had we not split the single long sequence into subsequences of about 1000 measurements, this would not have been a problem as the identification of a ``dark'' or ``bright'' result can always be done a posteriori based on the photon counts. However, 90 subsequences ended on a 5-count detection, from which a ``bright'' state was determined. In order to fix this, we mark those subsequences and all respectively following subsequences in which the first measurement was not along the right ray with a leading ``o'' in the data file, and omit them in the evaluation. For example, if subsequence $N$ ended on a 5-count detection and started with ray $v_{0,N}$, we mark the corresponding line in the data set and all following ones until and excluding the next line that starts with $v_{0,N}$. The new $N^\text{th}$ line can now again be correctly concatenated with the preceding one. While this reduces our dataset by around 1 million measurements, the impact on the numerical results is negligible.

%%%%%%%%%%%%%%%%%%%%%%%%%%%%%%%%%%%%%%%%%%%%%%%%%%%%%%%%%%%%%%%%%%%

\makeatletter
\iftoggle{arXiv}{\apptocmd{\thebibliography}{\global\c@NAT@ctr 35\relax}{}{}}{}
\makeatother

\iftoggle{arXiv}{\input{supp.bbl}}{\bibliography{myrefs,myextrarefs}}

%% file: fig_DetectionHistogram.tex
% !TeX root = testFigs.tex

%\pgfmathdeclarefunction{Poisson}{1}{\pgfmathparse{(#1^x)*exp(-#1)/(x!)}}

\pgfplotstableread{
x	y	yD	yB
0	17520942	1.750720262390098e7	0.12738546537061005
1	12383002	1.2411053380881079e7	2.388777649570045
2	4429881	4.399295528389759e6	22.3976049484289
3	1069106	1.0397580829679228e6	140.00262366156988
4	196775	184485.6445867066	656.3447747176897
5	32394	26379.110488091446	2461.60223020143
6	12410	3345.316081585575	7693.473731519368
7	22706	571.4825618099262	20610.108790601502
8	50861	285.94056062999204	48311.01324134548
9	104667	258.9580702591724	100660.59326597262
10	195904	254.17400749482258	188762.33225403237
11	328925	249.32746764676773	321794.41212849144
12	511167	241.96942662838896	502866.95922795
13	735099	231.36764551382578	725380.0347787818
14	977749	217.16708588169365	971613.210952533
15	1217917	199.4136664220636	1.2146691490523755e6
16	1419983	178.60586317419694	1.4236193010898347e6
17	1564097	155.65289437960664	1.5703656789885901e6
18	1629569	131.74032523449594	1.6360031624862582e6
19	1609836	108.13926987842706	1.6146797057853916e6
20	1508490	86.01038759979222	1.5139524674862863e6
21	1348758	66.24989692890051	1.351913155989065e6
22	1152631	49.40640159284909	1.1523435160114716e6
23	944075	35.67353198842539	939528.536193395
24	740602	24.943357890872534	734098.9157928706
25	557372	16.894702687956958	550643.3807083917
26	404950	11.089646580050402	397148.49754912936
27	283474	7.0578462823476915	275832.22882860026
28	190908	4.357638011304411	184732.29527063173
29	124214	2.611592786506011	119453.99460372973
30	78828	1.5201765549514477	74668.12942671482
31	48243	0.8599623725147731	45167.85084997354
32	28738	0.4730692828437969	26468.86368561115
33	16125	0.2532159231309202	15041.017148763376
34	9212	0.131957968095648	8295.721015237325
35	4951	0.06699015841434351	4444.694778642074
36	2686	0.033148492258337985	2315.236129850947
37	1440	0.01599683813773841	1173.4090028998369
38	685	0.007532784678870273	579.0574697572298
39	342	0.003463015641838898	278.42800157739134
40	139	0.0015550711544529473	130.529528083817
41	70	0.0006824256988690056	59.700883760492346
42	32	0.0002928027880570785	26.65552978489221
43	17	0.0001228876903805531	11.624511514761691
44	10	0.00005047157484494589	4.954249617105091
45	3	0.00002029439594786528	2.0645301027084293
46	1	7.99236747933857e-6	0.8416261810707927
47	0	3.084025822222243e-6	0.33579732172869997
48	1	1.1664635895507094e-6	0.13118731380206133
}{\dataDetectionHistogram}

\begin{tikzpicture}
\begin{axis}[width=6.5cm, height=4.5cm,
	ymode=log, xmin=-1, xmax=51, ymin=0.4, ymax=2e8, ytick distance=10^2,
	xlabel={PMT counts in \SI{160}{\micro\second}}, ylabel={Number of occurrences}]
%\addplot [thick, dotted, samples at = {0,...,10}, mark=none] {3.627e7*Poisson(0.709)};
%\addplot [thick, dashed, samples at = {0,...,50}, mark=none] {1.810e7*Poisson(18.751)};
\addplot [thick, dotted, mark=none] table [y=yD] {\dataDetectionHistogram};
\addplot [thick, dashed, mark=none] table [y=yB] {\dataDetectionHistogram};
\addplot [orange, only marks, mark=*, mark size=1.6pt] table {\dataDetectionHistogram};
\end{axis}
\end{tikzpicture}

%% file: fig_Repeatability.tex
% !TeX root = testFigs.tex

\begin{tikzpicture}
\begin{axis}[width=6.5cm, height=3cm,
	xmin=0.2, xmax=13.8, ymin=0.996, ymax=1,
	xtick pos=left, xtick=data, ytick pos=left,
	xticklabels={$y_1^-$, $y_2^-$, $y_3^-$, $y_1^+$, $y_2^+$, $y_3^+$, $h_1$,
			$h_2$, $h_3$, $h_0$, $z_1$, $z_2$, $z_3$},
	ylabel={Repeatability $R_{u}$},
	mark size=1pt,
	yticklabel style={/pgf/number format/fixed,
			/pgf/number format/fixed zerofill,
			/pgf/number format/precision=3}]
\addplot+[only marks, error bars/.cd, y dir=both, y explicit] 
	table[x=x,y=y,y error=error] {
		x	y			error
		1	0.9976983	0.000085
		2	0.9976227	0.000086
		3	0.9984565	0.000069
		4	0.9978012	0.000083
		5	0.9986974	0.000064
		6	0.9984901	0.000069
		7	0.9967358	0.000101
		8	0.9979029	0.000081
		9	0.9978861	0.000081
		10	0.9979026	0.000081
		11	0.9995397	0.000038
		12	0.9985860	0.000066
		13	0.9984671	0.000069
	};
\end{axis}
\end{tikzpicture}

%% file: fig_Signaling.tex
% !TeX root = testFigs.tex

%\pgfmathdeclarefunction{Poisson}{1}{\pgfmathparse{(#1^x)*exp(-#1)/(x!)}}

\pgfplotstableread{
x	y	fit
-5.75	0	0.2452
-5.25	3	0.6735
-4.75	1	1.6866
-4.25	8	3.8505
-3.75	11	8.014
-3.25	18	15.20
-2.75	22	26.30
-2.25	46	41.48
-1.75	59	59.65
-1.25	77	78.18
-0.75	89	93.43
-0.25	102	101.
0.25	104	101.1
0.75	91	91.547
1.25	82	75.572
1.75	53	56.874
2.25	39	39.022
2.75	17	24.408
3.25	14	13.919
3.75	13	7.2364
4.25	2	3.42979
4.75	3	1.48200
5.25	4	0.58381
5.75	0	0.20967
}{\dataSFo}

\pgfplotstableread{
x	y	fit
-5.75	0	0.
-5.25	0	0.00003
-4.75	0	0.00046
-4.25	0	0.00603
-3.75	0	0.05977
-3.25	0	0.44418
-2.75	3	2.47397
-2.25	10	10.326
-1.75	34	32.304
-1.25	74	75.736
-0.75	118	133.0
-0.25	199	175.2
0.25	171	172.90
0.75	103	127.87
1.25	88	70.8727
1.75	34	29.4384
2.25	17	9.16394
2.75	5	2.13788
3.25	2	0.373780
3.75	0	0.04898
4.25	0	0.00481
4.75	0	0.000350
5.25	0	0.00002
5.75	0	0.
}{\dataSBa}

\begin{tikzpicture}
\begin{axis}[width=6.5cm, height=4.5cm,
%	ymode=log, xmin=-1, xmax=51, ymin=0.4, ymax=2e8, ytick distance=10^2,
	minor x tick num=1, minor y tick num=4,
	mark size=1pt, legend style={draw=none, font=\small}, legend cell align=left,
	extra x ticks={0}, extra x tick labels=\empty, extra x tick style={grid=major},
	extra y ticks={0}, extra y tick labels=\empty, extra y tick style={grid=major},
	xlabel={$S_{u,v,w;i}^\text{(ba)}/\Delta S_{u,v,w;i}^\text{(ba)}$
		and $S_{u,v,w;i}^\text{(fo)}/\Delta S_{u,v,w;i}^\text{(fo)}$}, ylabel={Number of occurrences}]
\addplot [orange, thick, smooth, tension=0.6, mark=none] table [y=fit] {\dataSBa};
%	\addlegendentry{$\mu^\text{(ba)}$, $\sigma^\text{(ba)}$}
\addplot [blue, thick, smooth, tension=0.6, mark=none] table [y=fit] {\dataSFo};
%	\addlegendentry{$\mu^\text{(fo)}$, $\sigma^\text{(fo)}$}
\addplot [orange, only marks,] table [y=y] {\dataSBa};
\addplot [blue, only marks,] table [y=y] {\dataSFo};

\node [orange, align=left, font=\small] at (4,160)
	{$\begin{aligned}
	\mu^\text{(ba)}&=\num{-0.02(4)} \\ \sigma^\text{(ba)}&=\num{0.93(4)}
	\end{aligned}$};
\node [blue, align=left, font=\small] at (-4,130)
	{$\begin{aligned}
	\mu^\text{(fo)}&=\num{-0.04(4)} \\ \sigma^\text{(fo)}&=\num{1.64(4)}
	\end{aligned}$};

\end{axis}

\end{tikzpicture}

%% file: fig_Blind.tex
% !TeX root = testFigs.tex

\pgfplotsset{colormap={whiteblack}{gray=(1) color=(blue!50)}}
\pgfplotsset{matrixplot/.style={anchor=center, 
		width=4.5cm, height=4.5cm, enlargelimits=false, 
		xtick pos=right, tick style=transparent, typeset ticklabels with strut,
		ticklabel style={font=\footnotesize},
		xticklabels={$y_1^-$, $y_2^-$, $y_3^-$, $y_1^+$, $y_2^+$, $y_3^+$, 
			$h_1$, $h_2$, $h_3$, $h_0$, $z_1$, $z_2$, $z_3$},
		yticklabels={$y_1^-$, $y_2^-$, $y_3^-$, $y_1^+$, $y_2^+$, $y_3^+$, 
			$h_1$, $h_2$, $h_3$, $h_0$, $z_1$, $z_2$, $z_3$}}}

%\pgfplotstableread{
%
%}{\dataBlindDetermination}

\begin{tikzpicture} 

\def\ticklabels{1, 2, 3, 4, 5, 6, 7, 8, 9, 10, 11, 12, 13}
\begin{axis}[name=BlindIdeal, matrixplot,
	xtick=\ticklabels, ytick=\ticklabels]
	\addplot[matrix plot, mesh/cols=13, colormap name=whiteblack, point meta=explicit]
		table[meta=Ideal] {
			x	y	Ideal	Random	Ordered
			1	1	0	0	0
			2	1	0	0	0
			3	1	0	0	0
			4	1	1	0	1
			5	1	0	0	0
			6	1	0	1	0
			7	1	1	0	1
			8	1	0	0	0
			9	1	0	0	0
			10	1	1	1	1
			11	1	1	0	1
			12	1	0	1	0
			13	1	0	1	0
			1	2	0	0	0
			2	2	0	0	0
			3	2	0	0	0
			4	2	0	0	0
			5	2	1	1	1
			6	2	0	0	0
			7	2	0	1	0
			8	2	1	0	1
			9	2	0	0	0
			10	2	1	1	1
			11	2	0	0	0
			12	2	1	0	1
			13	2	0	0	0
			1	3	0	0	0
			2	3	0	0	0
			3	3	0	0	0
			4	3	0	0	0
			5	3	0	1	0
			6	3	1	0	1
			7	3	0	0	0
			8	3	0	1	0
			9	3	1	0	1
			10	3	1	0	1
			11	3	0	0	0
			12	3	0	0	0
			13	3	1	1	1
			1	4	1	0	1
			2	4	0	0	0
			3	4	0	0	0
			4	4	0	0	0
			5	4	0	0	0
			6	4	0	0	0
			7	4	0	0	0
			8	4	1	1	1
			9	4	1	0	1
			10	4	0	1	0
			11	4	1	1	1
			12	4	0	0	0
			13	4	0	0	0
			1	5	0	0	0
			2	5	1	1	1
			3	5	0	1	0
			4	5	0	0	0
			5	5	0	0	0
			6	5	0	1	0
			7	5	1	0	1
			8	5	0	0	0
			9	5	1	0	1
			10	5	0	0	0
			11	5	0	1	0
			12	5	1	0	1
			13	5	0	0	0
			1	6	0	1	0
			2	6	0	0	0
			3	6	1	0	1
			4	6	0	0	0
			5	6	0	1	0
			6	6	0	0	0
			7	6	1	0	1
			8	6	1	0	1
			9	6	0	0	0
			10	6	0	0	0
			11	6	0	1	0
			12	6	0	1	0
			13	6	1	0	1
			1	7	1	0	1
			2	7	0	1	0
			3	7	0	0	0
			4	7	0	0	0
			5	7	1	0	1
			6	7	1	0	1
			7	7	0	0	0
			8	7	0	1	0
			9	7	0	1	0
			10	7	0	0	0
			11	7	0	0	0
			12	7	0	1	0
			13	7	0	0	0
			1	8	0	0	0
			2	8	1	0	1
			3	8	0	1	0
			4	8	1	1	1
			5	8	0	0	0
			6	8	1	0	1
			7	8	0	1	0
			8	8	0	0	0
			9	8	0	0	0
			10	8	0	0	0
			11	8	0	0	0
			12	8	0	1	0
			13	8	0	0	0
			1	9	0	0	0
			2	9	0	0	0
			3	9	1	0	1
			4	9	1	0	1
			5	9	1	0	1
			6	9	0	0	0
			7	9	0	1	0
			8	9	0	0	0
			9	9	0	0	0
			10	9	0	0	0
			11	9	0	1	0
			12	9	0	0	0
			13	9	0	1	0
			1	10	1	1	1
			2	10	1	1	1
			3	10	1	0	1
			4	10	0	1	0
			5	10	0	0	0
			6	10	0	0	0
			7	10	0	0	0
			8	10	0	0	0
			9	10	0	0	0
			10	10	0	0	0
			11	10	0	0	0
			12	10	0	0	0
			13	10	0	1	0
			1	11	1	0	1
			2	11	0	0	0
			3	11	0	0	0
			4	11	1	1	1
			5	11	0	1	0
			6	11	0	1	0
			7	11	0	0	0
			8	11	0	0	0
			9	11	0	1	0
			10	11	0	0	0
			11	11	0	0	0
			12	11	1	0	1
			13	11	1	0	1
			1	12	0	1	0
			2	12	1	0	1
			3	12	0	0	0
			4	12	0	0	0
			5	12	1	0	1
			6	12	0	1	0
			7	12	0	1	0
			8	12	0	1	0
			9	12	0	0	0
			10	12	0	0	0
			11	12	1	0	1
			12	12	0	0	0
			13	12	1	0	1
			1	13	0	1	0
			2	13	0	0	0
			3	13	1	1	1
			4	13	0	0	0
			5	13	0	0	0
			6	13	1	0	1
			7	13	0	0	0
			8	13	0	0	0
			9	13	0	1	0
			10	13	0	1	0
			11	13	1	0	1
			12	13	1	0	1
			13	13	0	0	0
		};
\end{axis}
\node[left] at ($(BlindIdeal.outer north west)+(0.2,-0.3)$) {(a)};

\def\ticklabels{5, 10, 7, 11, 13, 8, 3, 4, 9, 2, 6, 1, 12}
\begin{axis}[name=BlindRandom, xshift=6cm, matrixplot,
	xtick={1,...,13}, xticklabels={a,b,c,d,e,f,g,h,i,j,k,l,m},
	ytick=\ticklabels]
	\addplot[matrix plot, mesh/cols=13, colormap name=whiteblack, point meta=explicit]
		table[meta=Random] {
			x	y	Ideal	Random	Ordered
			1	1	0	0	0
			2	1	0	0	0
			3	1	0	0	0
			4	1	1	0	1
			5	1	0	0	0
			6	1	0	1	0
			7	1	1	0	1
			8	1	0	0	0
			9	1	0	0	0
			10	1	1	1	1
			11	1	1	0	1
			12	1	0	1	0
			13	1	0	1	0
			1	2	0	0	0
			2	2	0	0	0
			3	2	0	0	0
			4	2	0	0	0
			5	2	1	1	1
			6	2	0	0	0
			7	2	0	1	0
			8	2	1	0	1
			9	2	0	0	0
			10	2	1	1	1
			11	2	0	0	0
			12	2	1	0	1
			13	2	0	0	0
			1	3	0	0	0
			2	3	0	0	0
			3	3	0	0	0
			4	3	0	0	0
			5	3	0	1	0
			6	3	1	0	1
			7	3	0	0	0
			8	3	0	1	0
			9	3	1	0	1
			10	3	1	0	1
			11	3	0	0	0
			12	3	0	0	0
			13	3	1	1	1
			1	4	1	0	1
			2	4	0	0	0
			3	4	0	0	0
			4	4	0	0	0
			5	4	0	0	0
			6	4	0	0	0
			7	4	0	0	0
			8	4	1	1	1
			9	4	1	0	1
			10	4	0	1	0
			11	4	1	1	1
			12	4	0	0	0
			13	4	0	0	0
			1	5	0	0	0
			2	5	1	1	1
			3	5	0	1	0
			4	5	0	0	0
			5	5	0	0	0
			6	5	0	1	0
			7	5	1	0	1
			8	5	0	0	0
			9	5	1	0	1
			10	5	0	0	0
			11	5	0	1	0
			12	5	1	0	1
			13	5	0	0	0
			1	6	0	1	0
			2	6	0	0	0
			3	6	1	0	1
			4	6	0	0	0
			5	6	0	1	0
			6	6	0	0	0
			7	6	1	0	1
			8	6	1	0	1
			9	6	0	0	0
			10	6	0	0	0
			11	6	0	1	0
			12	6	0	1	0
			13	6	1	0	1
			1	7	1	0	1
			2	7	0	1	0
			3	7	0	0	0
			4	7	0	0	0
			5	7	1	0	1
			6	7	1	0	1
			7	7	0	0	0
			8	7	0	1	0
			9	7	0	1	0
			10	7	0	0	0
			11	7	0	0	0
			12	7	0	1	0
			13	7	0	0	0
			1	8	0	0	0
			2	8	1	0	1
			3	8	0	1	0
			4	8	1	1	1
			5	8	0	0	0
			6	8	1	0	1
			7	8	0	1	0
			8	8	0	0	0
			9	8	0	0	0
			10	8	0	0	0
			11	8	0	0	0
			12	8	0	1	0
			13	8	0	0	0
			1	9	0	0	0
			2	9	0	0	0
			3	9	1	0	1
			4	9	1	0	1
			5	9	1	0	1
			6	9	0	0	0
			7	9	0	1	0
			8	9	0	0	0
			9	9	0	0	0
			10	9	0	0	0
			11	9	0	1	0
			12	9	0	0	0
			13	9	0	1	0
			1	10	1	1	1
			2	10	1	1	1
			3	10	1	0	1
			4	10	0	1	0
			5	10	0	0	0
			6	10	0	0	0
			7	10	0	0	0
			8	10	0	0	0
			9	10	0	0	0
			10	10	0	0	0
			11	10	0	0	0
			12	10	0	0	0
			13	10	0	1	0
			1	11	1	0	1
			2	11	0	0	0
			3	11	0	0	0
			4	11	1	1	1
			5	11	0	1	0
			6	11	0	1	0
			7	11	0	0	0
			8	11	0	0	0
			9	11	0	1	0
			10	11	0	0	0
			11	11	0	0	0
			12	11	1	0	1
			13	11	1	0	1
			1	12	0	1	0
			2	12	1	0	1
			3	12	0	0	0
			4	12	0	0	0
			5	12	1	0	1
			6	12	0	1	0
			7	12	0	1	0
			8	12	0	1	0
			9	12	0	0	0
			10	12	0	0	0
			11	12	1	0	1
			12	12	0	0	0
			13	12	1	0	1
			1	13	0	1	0
			2	13	0	0	0
			3	13	1	1	1
			4	13	0	0	0
			5	13	0	0	0
			6	13	1	0	1
			7	13	0	0	0
			8	13	0	0	0
			9	13	0	1	0
			10	13	0	1	0
			11	13	1	0	1
			12	13	1	0	1
			13	13	0	0	0
		};
\end{axis}
\node[left] at ($(BlindRandom.outer north west)+(0.2,-0.3)$) {(b)};

\def\ticklabels{5, 4, 3, 2, 1, 6, 7, 8, 10, 9, 12, 11, 13}
\begin{axis}[name=BlindSorted, xshift=12cm, matrixplot,
	xtick={11, 9, 7, 8, 5, 12, 3, 6, 10, 4, 2, 13, 1}, xticklabels={a,b,c,d,e,f,g,h,i,j,k,l,m},
	ytick=\ticklabels]
	\addplot[matrix plot, mesh/cols=13, colormap name=whiteblack, point meta=explicit]
		table[meta=Ordered] {
			x	y	Ideal	Random	Ordered
			1	1	0	0	0
			2	1	0	0	0
			3	1	0	0	0
			4	1	1	0	1
			5	1	0	0	0
			6	1	0	1	0
			7	1	1	0	1
			8	1	0	0	0
			9	1	0	0	0
			10	1	1	1	1
			11	1	1	0	1
			12	1	0	1	0
			13	1	0	1	0
			1	2	0	0	0
			2	2	0	0	0
			3	2	0	0	0
			4	2	0	0	0
			5	2	1	1	1
			6	2	0	0	0
			7	2	0	1	0
			8	2	1	0	1
			9	2	0	0	0
			10	2	1	1	1
			11	2	0	0	0
			12	2	1	0	1
			13	2	0	0	0
			1	3	0	0	0
			2	3	0	0	0
			3	3	0	0	0
			4	3	0	0	0
			5	3	0	1	0
			6	3	1	0	1
			7	3	0	0	0
			8	3	0	1	0
			9	3	1	0	1
			10	3	1	0	1
			11	3	0	0	0
			12	3	0	0	0
			13	3	1	1	1
			1	4	1	0	1
			2	4	0	0	0
			3	4	0	0	0
			4	4	0	0	0
			5	4	0	0	0
			6	4	0	0	0
			7	4	0	0	0
			8	4	1	1	1
			9	4	1	0	1
			10	4	0	1	0
			11	4	1	1	1
			12	4	0	0	0
			13	4	0	0	0
			1	5	0	0	0
			2	5	1	1	1
			3	5	0	1	0
			4	5	0	0	0
			5	5	0	0	0
			6	5	0	1	0
			7	5	1	0	1
			8	5	0	0	0
			9	5	1	0	1
			10	5	0	0	0
			11	5	0	1	0
			12	5	1	0	1
			13	5	0	0	0
			1	6	0	1	0
			2	6	0	0	0
			3	6	1	0	1
			4	6	0	0	0
			5	6	0	1	0
			6	6	0	0	0
			7	6	1	0	1
			8	6	1	0	1
			9	6	0	0	0
			10	6	0	0	0
			11	6	0	1	0
			12	6	0	1	0
			13	6	1	0	1
			1	7	1	0	1
			2	7	0	1	0
			3	7	0	0	0
			4	7	0	0	0
			5	7	1	0	1
			6	7	1	0	1
			7	7	0	0	0
			8	7	0	1	0
			9	7	0	1	0
			10	7	0	0	0
			11	7	0	0	0
			12	7	0	1	0
			13	7	0	0	0
			1	8	0	0	0
			2	8	1	0	1
			3	8	0	1	0
			4	8	1	1	1
			5	8	0	0	0
			6	8	1	0	1
			7	8	0	1	0
			8	8	0	0	0
			9	8	0	0	0
			10	8	0	0	0
			11	8	0	0	0
			12	8	0	1	0
			13	8	0	0	0
			1	9	0	0	0
			2	9	0	0	0
			3	9	1	0	1
			4	9	1	0	1
			5	9	1	0	1
			6	9	0	0	0
			7	9	0	1	0
			8	9	0	0	0
			9	9	0	0	0
			10	9	0	0	0
			11	9	0	1	0
			12	9	0	0	0
			13	9	0	1	0
			1	10	1	1	1
			2	10	1	1	1
			3	10	1	0	1
			4	10	0	1	0
			5	10	0	0	0
			6	10	0	0	0
			7	10	0	0	0
			8	10	0	0	0
			9	10	0	0	0
			10	10	0	0	0
			11	10	0	0	0
			12	10	0	0	0
			13	10	0	1	0
			1	11	1	0	1
			2	11	0	0	0
			3	11	0	0	0
			4	11	1	1	1
			5	11	0	1	0
			6	11	0	1	0
			7	11	0	0	0
			8	11	0	0	0
			9	11	0	1	0
			10	11	0	0	0
			11	11	0	0	0
			12	11	1	0	1
			13	11	1	0	1
			1	12	0	1	0
			2	12	1	0	1
			3	12	0	0	0
			4	12	0	0	0
			5	12	1	0	1
			6	12	0	1	0
			7	12	0	1	0
			8	12	0	1	0
			9	12	0	0	0
			10	12	0	0	0
			11	12	1	0	1
			12	12	0	0	0
			13	12	1	0	1
			1	13	0	1	0
			2	13	0	0	0
			3	13	1	1	1
			4	13	0	0	0
			5	13	0	0	0
			6	13	1	0	1
			7	13	0	0	0
			8	13	0	0	0
			9	13	0	1	0
			10	13	0	1	0
			11	13	1	0	1
			12	13	1	0	1
			13	13	0	0	0
		};
\end{axis}
\node[left] at ($(BlindSorted.outer north west)+(0.2,-0.3)$) {(c)};

\end{tikzpicture}

%% file: fig_YOSh.tex
\pgfplotsset{YOSh/.style={
	width=25mm, y=3.7mm, scale only axis, clip=false,
	xtick pos=left, ytick pos=left, ytick=data,
	visualization depends on={value \thisrow{label} \as \labela},
	mark size=1pt, nodes near coords align={horizontal}}}

\begin{tikzpicture}[every node/.style={font=\footnotesize}]
\begin{axis}[YOSh, name=Opt3Sh,
	xmin=25, xmax=28.2, ymin=-13.8,	ymax=-0.4,
	yticklabels={$\expval*{\chi_\text{opt3}}_{y_1^-}$, $\expval*{\chi_\text{opt3}}_{y_2^-}$, $\expval*{\chi_\text{opt3}}_{y_3^-}$, $\expval*{\chi_\text{opt3}}_{y_1^+}$, $\expval*{\chi_\text{opt3}}_{y_2^+}$, $\expval*{\chi_\text{opt3}}_{y_3^+}$, $\expval*{\chi_\text{opt3}}_{h_1}$, $\expval*{\chi_\text{opt3}}_{h_2}$, $\expval*{\chi_\text{opt3}}_{h_3}$, $\expval*{\chi_\text{opt3}}_{h_0}$, $\expval*{\chi_\text{opt3}}_{z_1}$, $\expval*{\chi_\text{opt3}}_{z_2}$, $\expval*{\chi_\text{opt3}}_{z_3}$}]
\draw[dashed] (27.6667,\pgfkeysvalueof{/pgfplots/ymin})
	-- (27.6667,\pgfkeysvalueof{/pgfplots/ymax});
\draw[Stealth-] (27.6667,\pgfkeysvalueof{/pgfplots/ymax}) |- ++(-4mm,3mm) node[anchor=east] {27.67};
\draw[Stealth-] (25,\pgfkeysvalueof{/pgfplots/ymax}) |- ++(4mm,3mm) node[anchor=west] {25};
\addplot+[only marks, error bars/.cd, x dir=both, x explicit] 
	table[x=x,y=y,x error=error] {
		x		error	y	label
		27.3974	0.051	-1	\num{27.40(6)}
		27.2505	0.051	-2	\num{27.25(6)}
		27.3037	0.052	-3	\num{27.30(6)}
		27.4308	0.051	-4	\num{27.43(6)}
		27.3693	0.051	-5	\num{27.37(6)}
		27.5474	0.051	-6	\num{27.55(6)}
		27.2539	0.052	-7	\num{27.25(6)}
		27.3040	0.052	-8	\num{27.30(6)}
		27.4648	0.051	-9	\num{27.46(6)}
		27.5172	0.051	-10	\num{27.52(6)}
		27.4035	0.051	-11	\num{27.40(6)}
		27.4218	0.051	-12	\num{27.42(6)}
		27.3088	0.051	-13	\num{27.31(6)}
	};
\addplot [nodes near coords={\labela}, only marks, mark=none] 
	table[x expr=27, y=y] {
		x		error	y	label
		27.3974	0.051	-1	\num{27.40(6)}
		27.2505	0.051	-2	\num{27.25(6)}
		27.3037	0.052	-3	\num{27.30(6)}
		27.4308	0.051	-4	\num{27.43(6)}
		27.3693	0.051	-5	\num{27.37(6)}
		27.5474	0.051	-6	\num{27.55(6)}
		27.2539	0.052	-7	\num{27.25(6)}
		27.3040	0.052	-8	\num{27.30(6)}
		27.4648	0.051	-9	\num{27.46(6)}
		27.5172	0.051	-10	\num{27.52(6)}
		27.4035	0.051	-11	\num{27.40(6)}
		27.4218	0.051	-12	\num{27.42(6)}
		27.3088	0.051	-13	\num{27.31(6)}
	};
\end{axis}

\begin{axis}[YOSh, name=YOSh, 
	at={(Opt3Sh.north west)}, xshift=-19mm, anchor=north east,
	xmin=8, xmax=8.44,  ymin=-13.8,	ymax=-0.4,
	yticklabels={$\expval*{\chi_\text{YO}}_{y_1^-}$, $\expval*{\chi_\text{YO}}_{y_2^-}$, $\expval*{\chi_\text{YO}}_{y_3^-}$, $\expval*{\chi_\text{YO}}_{y_1^+}$, $\expval*{\chi_\text{YO}}_{y_2^+}$, $\expval*{\chi_\text{YO}}_{y_3^+}$, $\expval*{\chi_\text{YO}}_{h_1}$, $\expval*{\chi_\text{YO}}_{h_2}$, $\expval*{\chi_\text{YO}}_{h_3}$, $\expval*{\chi_\text{YO}}_{h_0}$, $\expval*{\chi_\text{YO}}_{z_1}$, $\expval*{\chi_\text{YO}}_{z_2}$, $\expval*{\chi_\text{YO}}_{z_3}$},
	xticklabel style={/pgf/number format/fixed, /pgf/number format/fixed zerofill, /pgf/number format/precision=1}]
\draw[dashed] (8.3333,\pgfkeysvalueof{/pgfplots/ymin}) -- (8.3333,\pgfkeysvalueof{/pgfplots/ymax});
\draw[Stealth-] (8.3333,\pgfkeysvalueof{/pgfplots/ymax}) |- ++(-4mm,3mm) node[anchor=east] {8.33};
\draw[Stealth-] (8,\pgfkeysvalueof{/pgfplots/ymax}) |- ++(4mm,3mm) node[anchor=west] {8};
\addplot+[only marks, error bars/.cd, x dir=both, x explicit] 	
	table[x=x,y=y,x error=error] {
		x		error	y	label
		8.2891	0.0164	-1	\num{8.29(2)}
		8.2790	0.0164	-2	\num{8.28(2)}
		8.3045	0.0164	-3	\num{8.30(2)}
		8.2794	0.0164	-4	\num{8.28(2)}
		8.2539	0.0164	-5	\num{8.25(2)}
		8.2891	0.0164	-6	\num{8.29(2)}
		8.2794	0.0165	-7	\num{8.28(2)}
		8.2809	0.0165	-8	\num{8.28(2)}
		8.2980	0.0164	-9	\num{8.30(2)}
		8.2644	0.0165	-10	\num{8.26(2)}
		8.2763	0.0161	-11	\num{8.28(2)}
		8.2878	0.0161	-12	\num{8.29(2)}
		8.2501	0.0161	-13	\num{8.25(2)}
	};
\addplot [nodes near coords={\labela}, only marks, mark=none] 
	table[x expr=8.23, y=y] {
		x		error	y	label
		8.2891	0.0164	-1	\num{8.29(2)}
		8.2790	0.0164	-2	\num{8.28(2)}
		8.3045	0.0164	-3	\num{8.30(2)}
		8.2794	0.0164	-4	\num{8.28(2)}
		8.2539	0.0164	-5	\num{8.25(2)}
		8.2891	0.0164	-6	\num{8.29(2)}
		8.2794	0.0165	-7	\num{8.28(2)}
		8.2809	0.0165	-8	\num{8.28(2)}
		8.2980	0.0164	-9	\num{8.30(2)}
		8.2644	0.0165	-10	\num{8.26(2)}
		8.2763	0.0161	-11	\num{8.28(2)}
		8.2878	0.0161	-12	\num{8.29(2)}
		8.2501	0.0161	-13	\num{8.25(2)}
	};
\end{axis}

\node [font={\small}, anchor=west] at ($(YOSh.outer north west)+(0mm,-1mm)$)	{(a)};
\node [font={\small}, anchor=west] at ($(Opt3Sh.outer north west)+(0mm,-1mm)$)	{(b)};

%\draw  (YOSh.outer south west) rectangle ++(8.6,6);

\end{tikzpicture}